\documentclass[nonacm]{acmart}

\pdfoutput=1

\AtBeginDocument{%
  }

\setcopyright{none}
\copyrightyear{2022}
\acmYear{2022}
\acmDOI{XXXXXXX.XXXXXXX}

\acmConference[CHI]{Make sure to enter the correct
  conference title from your rights confirmation emai}{June 03--05,
  2018}{Woodstock, NY}
\acmPrice{15.00}
\acmISBN{978-1-4503-XXXX-X/18/06}

\usepackage{hyperref}       %
\usepackage{url}            %
\usepackage{listings}
\usepackage{pdfpages}

\usepackage{todonotes}

\definecolor{codegreen}{rgb}{0,0.6,0}
\definecolor{codegray}{rgb}{0.5,0.5,0.5}
\definecolor{codepurple}{rgb}{0.58,0,0.82}
\definecolor{backcolour}{rgb}{0.95,0.95,0.92}
\lstdefinestyle{mystyle}{
    backgroundcolor=\color{backcolour},   
    commentstyle=\color{codegreen},
    keywordstyle=\color{magenta},
    numberstyle=\tiny\color{codegray},
    stringstyle=\color{codepurple},
    basicstyle=\ttfamily\footnotesize,
    breakatwhitespace=false,         
    breaklines=true,                 
    captionpos=b,                    
    keepspaces=true,                 
    numbers=none,                    
    numbersep=2pt,                  
    showspaces=false,                
    showstringspaces=false,
    showtabs=false,                  
    tabsize=2,
    breakindent=0pt,
}
\begin{document}

\title{Co-Writing Screenplays and Theatre Scripts with Language Models \\ An Evaluation by Industry Professionals}

\author{Piotr Mirowski}
\authornote{Both authors contributed equally to this research.}
\email{piotrmirowski@deepmind.com}
\author{Kory W. Mathewson}
\authornotemark[1]
\affiliation{%
  \institution{DeepMind}
  \country{United Kingdom}
}

\author{Jaylen Pittman}
\authornote{Work done while at DeepMind}
\affiliation{
  \institution{Stanford University}
  \country{USA}
}
\email{jaylenp@deepmind.com}

\author{Richard Evans}
\affiliation{%
  \institution{DeepMind}
  \country{United Kingdom}
}
\email{richardevans@deepmind.com}

\renewcommand{\shortauthors}{Mirowski and Mathewson et al.}

\begin{abstract}
Language models are increasingly attracting interest from writers. However, such models lack long-range semantic coherence, limiting their usefulness for longform creative writing. We address this limitation by applying language models hierarchically, in a system we call \textbf{Dramatron}. By building structural context via prompt chaining, Dramatron can generate coherent scripts and screenplays complete with title, characters, story beats, location descriptions, and dialogue. We illustrate Dramatron's usefulness as an interactive co-creative system with a user study of $15$ theatre and film industry professionals. Participants co-wrote theatre scripts and screenplays with Dramatron and engaged in open-ended interviews. We report critical reflections both from our interviewees and from independent reviewers who watched stagings of the works to illustrate how both Dramatron and hierarchical text generation could be useful for human-machine co-creativity. Finally, we discuss the suitability of Dramatron for co-creativity, ethical considerations---including plagiarism and bias---and participatory models for the design and deployment of such tools.

\end{abstract}

\maketitle

\section{Introduction}
\label{sec:introduction}
Large language models (LLMs) become  remarkable and useful in co-creative applications as their ability to generate text improves ~\citep{brown2020language,keskar2019ctrl,DBLP:journals/corr/abs-1912-02164}. 
While their use is primarily limited to assisting in natural language processing tasks~\citep{devlin2018bert,Wang2018GLUEAM}, these models show particular promise for automatic story generation~\citep{alabdulkarim2021automatic,reif2021recipe} as an augmentative tool for human writers~\citep{swanson-etal-2021-story} and for live performance.
Examples of such creative uses of LLMs include the generation of the script of short film \emph{Sunspring} (2016)~\cite{sunspring2016}, \emph{It's No Game} (2017), \emph{Sollicitors} (2020)~\cite{sollicitors2020} or \emph{The First Horror Movie Written Entirely By Bots} (2021)~\cite{thefirsthorror2021},
improvisational theatre alongside robots by company \emph{Improbotics} (2016)~\citep{mathewson2017improvised,mathewson2017turing,mathewson2018improbotics,branch2021collaborative}, collaborative script writing for theatre play \emph{AI}~\cite{ai2021tang}, and \emph{THEaiTRE} company's~\cite{rosa2020theaitre,rosa2021theaitre,rosa2022gpt,schmidtova2022dialoguescript} \emph{AI: When a Robot Writes a Play} (2021).

Models able to generate \emph{coherent} stories could be useful for co-writing theatre scripts and screenplays. This is a difficult task for LLMs because the narrative of a script or screenplay must exhibit long-term coherence and reincorporation, and LLMs are limited in their ability to model long-range dependencies (e.g., to reincorporate information from many pages ago). 
This limitation stems from the context window of LLMs, which today is limited to at most $2048$ tokens (i.e. about $1500$ words) in state-of-the-art models~\citep{gopher,openai_2021}.

\begin{figure}[t!]
    \centering
    \includegraphics[height=1.8in]{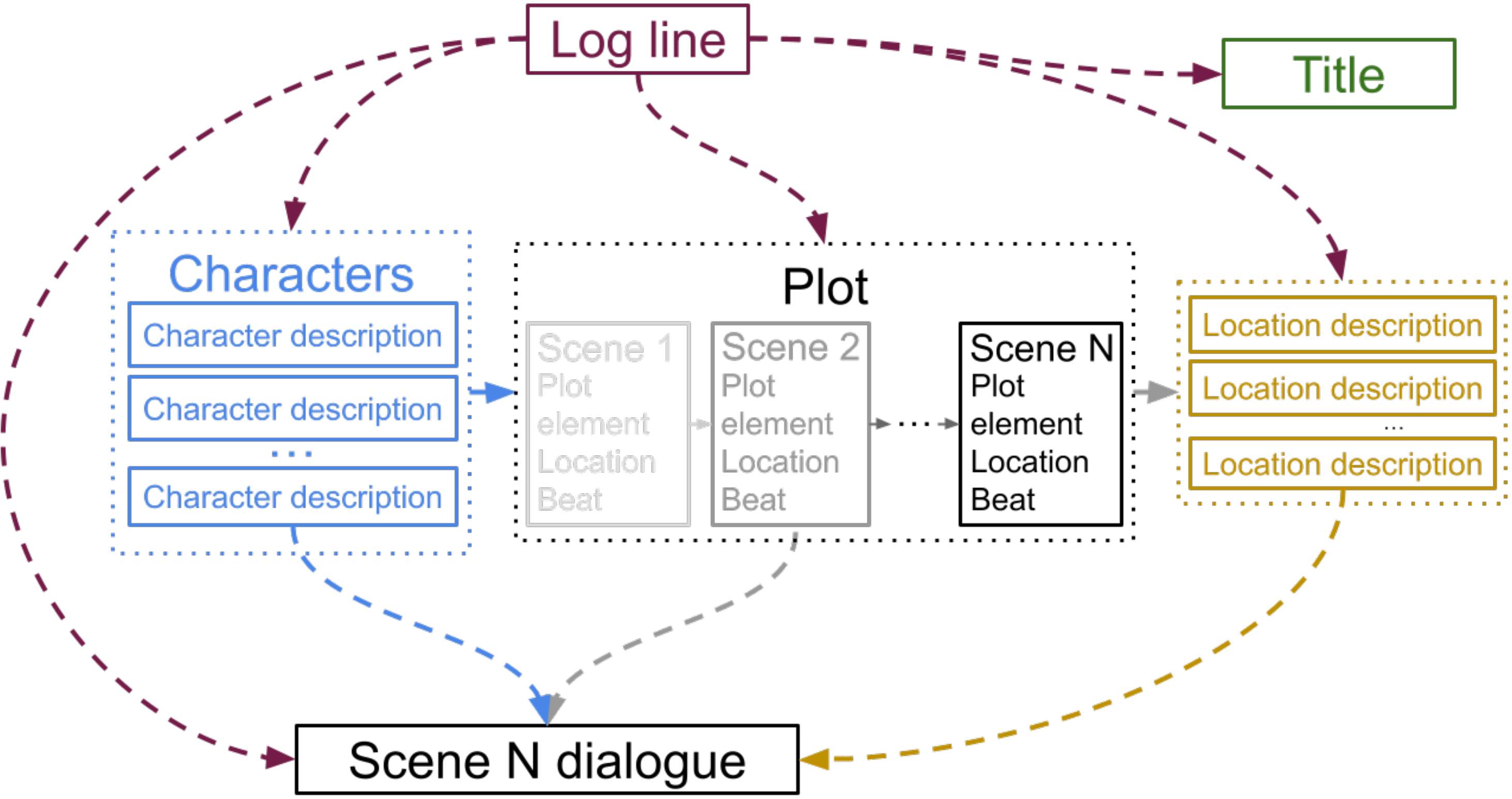}
    \caption{Dramatron's Hierarchical Coherent Story Generation. Dramatron starts from a log line to generate a title and characters. Characters generated are used as prompts to generate a sequence of scene summaries in the plot. Descriptions are subsequently generated for each unique location. Finally, these elements are all combined to generate dialogue for each scene. The arrows in the figure indicate how text generated is used to construct prompts for further LLM text generation.}
    \label{fig:dramatron-diagram}
\end{figure}

In this work, we present {\bf Dramatron}, a system that uses LLMs to generate scripts and screenplays hierarchically through a method we call \emph{hierarchical story generation}. Dramatron leverages the strengths of LLMs and combines well-designed prompts (see Appendix \ref{prompts}) and prompt chaining~\citep{wu2022ai} with structured generation for long range coherence across the entire script. This process results in greater story coherence than ``flat'' sequential text generation. 
Our method is, in spirit, similar to hierarchical neural story generation~\citep{fan2018hierarchical}, but generates scripts that far surpass $1000$ words. Hierarchical generation of stories can produce an entire script---sometimes tens of thousands of words---from a single user-provided summary of the central dramatic conflict, called the \emph{log line}~\citep{steiff2005complete}. 
From the input log line, Dramatron can generate an entire script with a title, list of characters, a plot (i.e. a list of scene summaries with settings and beats), location descriptions, and dialogue (see Fig.~\ref{fig:dramatron-diagram}). 
The user can intervene at any stage of the hierarchical generation. They can solicit alternative generations, edit and rewrite output text, or continue text generation. In this way, the human interactively co-writes the script.
Our methods can be used with any LLMs that accept an input prompt and then predict which tokens come next.

To evaluate Dramatron's usability and capabilities, instead of relying on online crowd-sourced annotation and evaluation from non-expert raters, we engaged $15$-experts in two-hour long user study sessions to co-write a script alongside Dramatron. 
The experts playwrights and screenwriters from the theatre and film industry were paid a consulting fee for their engagement.
They provided feedback on both the interactive co-authorship process, and artistic opinion and analysis of the outputs co-written with Dramatron.
Our inclusive research methodology invited participation from experts during the creative design and development process: their feedback directly led to incremental improvements of the system.
We provide a summary of the iterative tool refinement process that emerged from their feedback.
A collection of scripts co-written with this process were produced and staged at the Edmonton International Fringe Theatre Festival in August 2022. 
Reflections from the creative team are presented, as are comments from reviewers, as these represent critical reflections on human-machine co-creativity.
Our study design and data collection process was validated and approved by HuBREC (Human Behavioral Research Ethics Committee), which is a research ethics committee run within Deepmind which includes and is chaired by academics from outside the company.
To the best of our knowledge, this work represents the largest expert user study conducted on co-creative authorship to date.

The paper is structured as follows.
Section \ref{sec:storytelling} provides background on storytelling and how log lines become full length scripts.
Section \ref{sec:script-models} provides background on LLMs and their use in creative text generation and details on interaction with Dramatron.
Section \ref{sec:methods} provides details on the design of our human co-authorship study.
Section \ref{sec:results-interviews} presents the major themes summarizing the qualitative interviews. 
Section \ref{sec:results-surveys} covers the quantitative results from the human user study. 
Section \ref{sec:disc-and-conclusion} explores the potential impact of these systems on the broader creative community.
Finally, the Appendix includes related work on automated story generation (Appendix~\ref{sec:related}), as well as detailed prompt sets (Appendix~\ref{prompts}), an example of a raw generated script (Appendix~\ref{generated-output-script}) and four examples of edited scripts (Appendix~\ref{edited-output-script}).
Overall, this paper presents \textbf{Dramatron} and a pathway toward human-machine co-creativity that uplifts human writers and artists while leveraging novel artificial intelligence systems such as LLMs.

\section{Storytelling, The Shape of Stories, and Log Lines}
\label{sec:storytelling}

\begin{figure}[t!]
    \centering
    \includegraphics[width=0.4\linewidth]{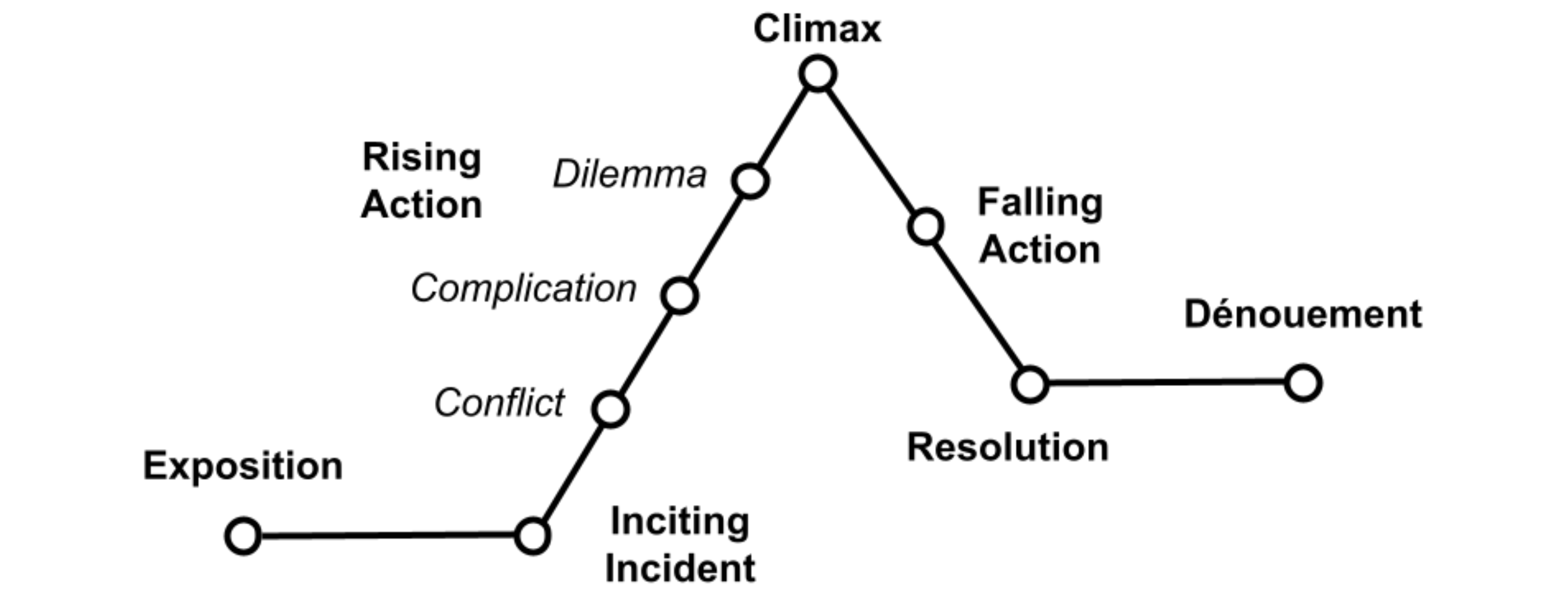}
    \includegraphics[width=0.35\linewidth]{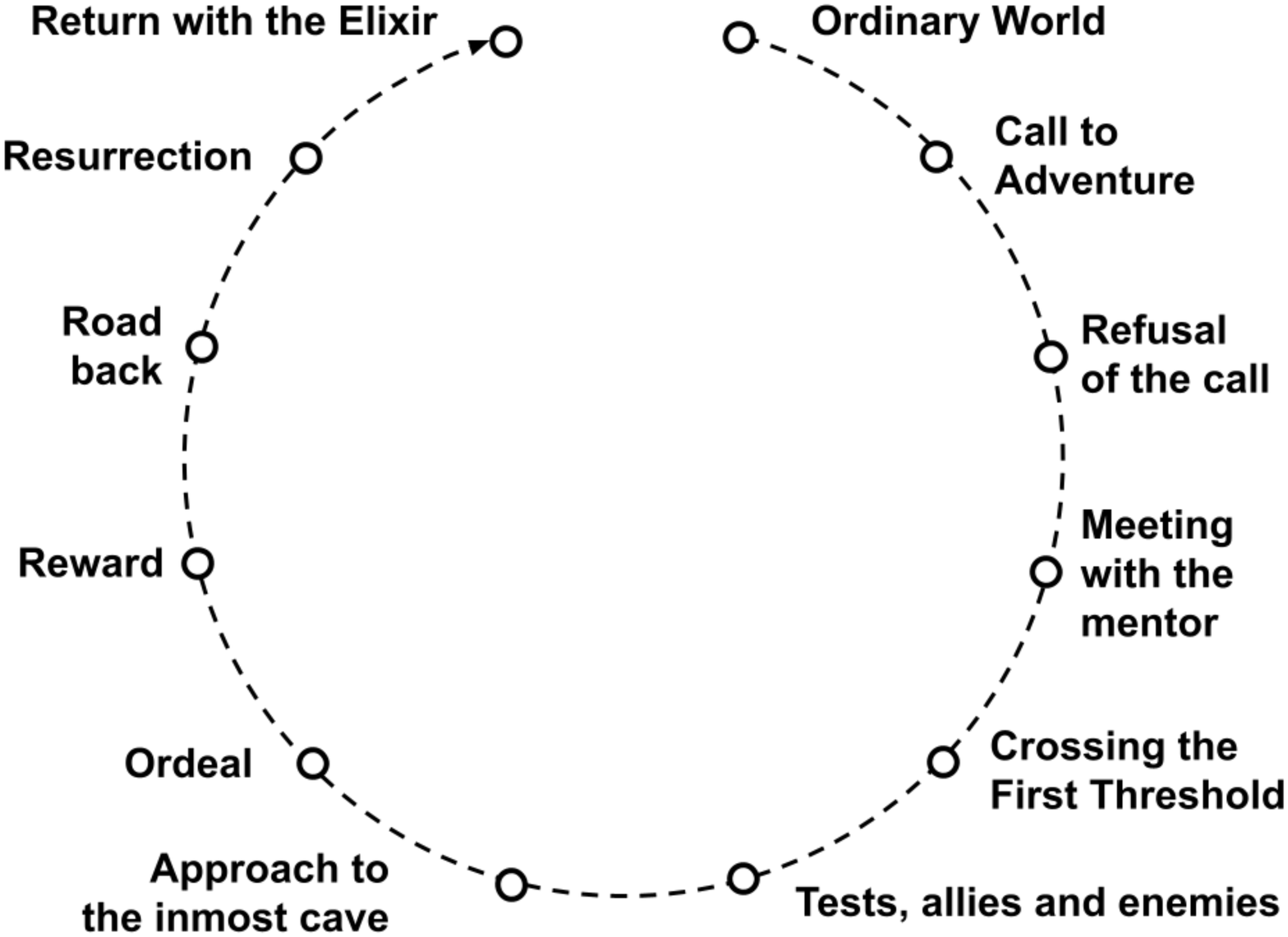}
    \setlength{\belowcaptionskip}{-20pt}
    \caption{Illustrations of two narrative structures employed in this study. Left: Gustav Freytag's pyramid. Right: Hero's Journey or \emph{Monomyth} using  \citet{vogler2007writer} formulation.}
    \label{fig:narrative-structure}
\end{figure}

Archaeological evidence including artists' depictions of narratives on jars from as far back as $2400$ BCE, show how ancient societies used the technologies of their time to help tell their stories~\citep{Ghosh1990AnEO}.
In \emph{Poetics}~\citep{aristotle350BCpoetics}, Aristotle identified plot (or mythos) as the most important part of a tragic story---the other elements being characters, their thoughts, language, melody and spectacle.
The plot is the sequence of actions that shape the story. 
Each plot point is coherent with, and a consequence of, the previous point(s). 
Aristotle introduced the well-known simple $3$ point plot: \textit{beginning}, \textit{middle}, and \textit{end}. 
There are many dramatic structures extending the work of Aristotle~\citep{polti1917thirty}.
For example, one of the many adaptations of Freytag's pyramid~\citep{freytag1894technik}, popular in Western storytelling is illustrated in Fig. \ref{fig:narrative-structure}. It includes this sequence of plot points: \emph{Exposition}, \emph{Inciting Incident}, a \emph{Rising Action} composed of a series of \emph{Conflicts}, \emph{Complications}, and \emph{Dilemmas}, \emph{Climax}, \emph{Falling Action}, \emph{Resolution}, and \emph{Dénouement}.
Seminal work in narrative analysis~\cite{labov1967narrative, propp1968morphology, rumelhart1975notes} suggests general but mutable structures present in storytelling processes across many linguistic and cultural traditions (e.g. abstract, orientation, complicating action, and coda narrative clauses). The finding that narratives in many societies and in various languages make use of large, modular elements has inspired the use of models of narrative in automated storytelling, as can be seen in fields such as computational narratology~\citep{johnstone2005discourse, mani2012computational}. The general structures found in the Structuralist narratology of Propp~\citep{propp1968morphology} and the Personal Experience Narratives of Labov and Waletzky~\citep{labov1967narrative} are aligned with Freytag's pyramid, which we choose because it is more in line with the specific discourse genre of dramatic scripts, and arguably more familiar to the playwrights we engaged with over the course of our study. However, we note that our choice of narrative structure is not universal in scope, and is, in fact, "characteristically Western"~\citep{johnstone2005discourse, de2015discourse}. Alternative story shapes, or "story grammars"~\citep{rumelhart1980evaluating} are possible~\citep{becker1979text, chafe1980pear}.
Fig. \ref{fig:narrative-structure} also shows an alternative story structure: the Hero's Journey or \emph{Monomyth}~\citep{campbell2008hero,vogler2007writer}. For narrative coherence, we leverage plot beats in our hierarchical generation.

Oftentimes the seed of narrative inspiration for a screenplay or theatre script is a log line~\citep{logline}. 
The log line summarizes the story in a few sentences and
is often adapted over the course of production due to creative team preferences and cultural references.
Log lines typically contain the setting, protagonist, antagonist, a conflict or goal, and sometimes the inciting incident~\citep{bishop2016story}. 
In this work, we use log lines to start the hierarchical story generation process because it contains plot elements in the answers to questions: \emph{Who? What? When and Where? How? Why?}~\citep{shimmin_2021}

\section{The Use of Large Language Models for Creative Text Generation}
\label{sec:script-models}
\subsection{Language Models}
\label{sec:chinchilla}

\begin{figure}[t!]
    \centering
    \includegraphics[width=0.85\linewidth]{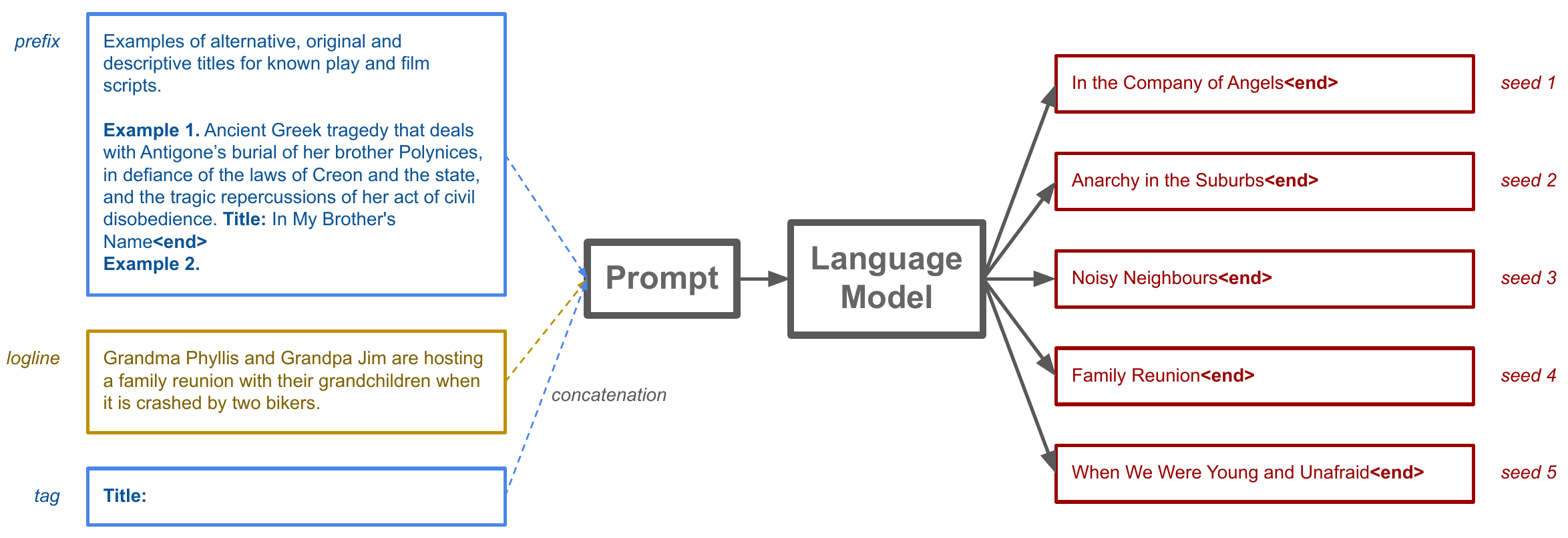}
    \caption{Illustration of the prompting setup for the language model, with user- or Dramatron-generated prompt being concatenated to a \emph{prefix} and decorated with \emph{tags}. Several title outputs are generated for different random seeds, each ending with \emph{tag} {\bf $<$end$>$}.}
    \label{fig:language-model}
\end{figure}

Statistical language models (language models, or LMs) model the probability of text \emph{tokens} given a \emph{context} of previous tokens---tokens can be words, characters, or character bi-grams.
Using machine learning, LMs are trained on large corpora of text to approximate the conditional probability distribution. 
LMs can compute the likelihood of a piece of text and generate new text as the continuation of a text \emph{prompt}. 
Text generation is probabilistic and involves random \emph{sampling} from the conditional probabilities. 
Different random seeds result in different random samples.
Figure~\ref{fig:language-model} illustrates an example of feeding a text prompt and using the LM to generate different text samples. 

In this study, we employed the \emph{Chinchilla} large language model (LLM)~\citep{hoffmann2022training}, represented as a neural network with $70$B-parameters and that was trained on $1.4$T tokens of the \emph{MassiveText} dataset. 
As described by~\citet{gopher}, that corpora contains $604$M \emph{MassiveWeb} documents, $4$M \emph{Books}, $361$M questions and responses from \emph{C4}, $1.1$B \emph{News} articles, $142$M \emph{GitHub} code entries, and $6$M \emph{Wikipedia} articles. 
Note that alternative LLMs could be employed, such as GPT-3.\footnote{Accessible at: \url{https://openai.com/api/}}

\subsection{Hierarchical Language Generation to Circumvent Limited Contexts}
\label{sec:hierarchical}

LLMs give the impression of coherence within and between paragraphs~\citep{bender2021dangers}, but have difficulty with long-term semantic coherence due to the restricted size of their context windows.
Memory wise, they require $\mathcal{O}(n^2)$ (where $n$ is the number of tokens in the context window). Thus, these models currently restrict $n$ to $2048$ tokens \citep{brown2020language,openai_2021}.
Because these models do not achieve long-term semantic coherence, the current state-of-the-art method for longer text generation is incorporating a human-in-the-loop who selects from various generations sampled from the model~\citep{mathewson2018improbotics,mirowski2019human,branch2021collaborative}. 
The human is tasked with achieving long-term semantic coherence: the model generates choices for the next textual snippet, while the human does the selection.\footnote{For example: \url{ https://theguardian.com/commentisfree/2020/sep/08/robot-wrote-this-article-gpt-3}}

In this project, we desire a system that can generate an entire text exhibiting long-term semantic coherence without necessarily requiring a human-in-the-loop. 
We \emph{encourage} writers to edit and modify the script at every level of the hierarchy. 
But, we do not \emph{require} a human-in-the-loop to achieve long-term semantic coherence. 
The hierarchical method will generate an entire script exhibiting reasonable long-term coherence from a single log line without human intervention.
Our approach to achieve long-term semantic coherence is to generate the story \emph{hierarchically}.

Our narrative generation is divided into $3$ hierarchical layers of abstraction.
The highest layer is the log line defined in Section~\ref{sec:storytelling}: a single sentence describing the central dramatic conflict.
The middle layer contains character descriptions, a plot outline (a sequence of high-level scene descriptions together with corresponding locations), and location descriptions.
The bottom layer is the actual character dialogue for the text of the script.
In this way, content at each layer is coherent with content in other layers. 
Note that ``coherent'' here refers to ``forming a unified whole'', not assuming any common sense and logical or emotion consistency to the LLM-generated text.

As illustrated on Figure~\ref{fig:dramatron-diagram}, the story is generated top-down \citep{rumelhart1975notes,thorndyke1977cognitive,wardrip2009expressive}.
After the human provides the log line,
Dramatron generates a list of characters, then a plot, and then descriptions of each location mentioned in the plot. 
Characters, plot, and location descriptions all meet the specification in the log line, in addition to causal dependencies, enabled by prompt chaining~\citep{wu2022ai} and explained on the diagram of Figure~\ref{fig:dramatron-diagram}.
Finally, for each scene in the plot outline, Dramatron generates dialogue satisfying previously generated scene specifications.
Resulting dialogues are appended together to generate the final output.
This hierarchical generation was designed to enable long-term semantic coherence.
A similar albeit inverted, method of recursive task decomposition was used to generate plot summaries~\citep{wu2021recursively}.
The incorporation of the middle layer, where the plot is summarised as a sequence of abstract scene descriptions, allows the entire plot to fit within the language models' context window. 
This overcomes prior limitations on long-term semantic coherence. 
Our method makes it possible for elements in the final scene to provide dramatic closure on elements introduced in the opening scene\footnote{See e.g. Chekhov's gun \citep{debreczeny1984chekhov}.}, and for generated stories to follow narrative arcs (see Section~\ref{sec:storytelling}).

\subsection{The Importance of Prompt Engineering}
\label{sec:prompts}

Dramatron uses several hard-coded prompts (i.e. input prefixes) to guide the large language model. 
Prompt engineering is a common way that users control or influence LLMs~\citep{brown2020language}. 
Each prompt has a few examples of desirable outputs. 
These are included in the prefix and adaptation to only a handful of examples is sometimes referred to as few-shot learning.
As illustrated in Figure~\ref{fig:language-model}, prompts are concatenated with user-supplied inputs and/or outputs of previous LLM generations.
This method is called prompt chaining~\citep{wu2022ai}, which is a type of algorithmic prompting~\citep{creswell2022faithful}. 
At lower levels of the hierarchy (see Fig.~\ref{fig:dramatron-diagram}), prompts are chained together with outputs from higher levels of the hierarchy.

In this work, we primarily used two sets of prompts: one based on Ancient Greek tragedy \emph{Medea} by Euripides, and one based on science-fiction films. For Dramatron, each prompt set is composed of: 1) title prompt, 2) character description prompt, 3) plot prompt, 4) location description prompt, 5) and dialogue prompt. 
Each prompt is detailed briefly below to give a sense of how they are engineered; additional details are in Appendix~\ref{prompts}. 

The \textbf{Title Prompt} is used to generate titles from a log line. A simplified title prompt, a user-provided log line, and randomly sampled titles are shown in Figure~\ref{fig:language-model}. 
It shows a prefix with an instruction (\texttt{Examples of alternative, original and descriptive titles for known play and film scripts.}) and an example (\texttt{Example 1. Ancient Greek tragedy [...]. Title: In My Brother's Name<end>}). The prefix finishes with: \texttt{Example 2.}
A user-input log line (e.g., \texttt{Grandma Phyllis and Grandpa Jim [...]}) is concatenated to that prefix, as well as the tag \texttt{Title:}, which encourages the LLM to generate a title that matches the log line. 
From a few examples, the LLM has ``learned'' to generate a related title and terminate tag \texttt{<end>}. The \textbf{Character Description Prompt} is used to generate character names and descriptions from a log line. The \textbf{Plot Outline Prompt} is used to turn a log line and list of characters into a plot. This prompt encourages the few-shot language model to transform a single sentence log line into a sequence of scene descriptions. 
Each scene is highly compressed, describing only the short name of the location, the narrative element identifying the position of the scene in the narrative arc (see Sec. \ref{sec:storytelling}), and a summary of what the characters are doing and saying, often called a narrative \emph{beat}\citep{mckee1997story}.
As a note, the prompt imposes a strong representational constraint on the way Dramatron represents a scene;
each scene is composed of a location, narrative element identifier, and beat. The \textbf{Location Description Prompt} is used to generate a detailed scenic description from a place name and a log line. Finally, the \textbf{Dialogue Prompt} is used to turn a beat (i.e., the scene summary), scene location description, description of each of the characters involved in the scene, and the log line (for story consistency), into dialogue. This prompt uses scene information generated for both the current and previous scenes.

\subsection{Interactive Writing with Dramatron}

As described above, with just few-shot (i.e., $1$-$4$) prompts and the user input log line, we leverage trained LLMs to generate complete scripts and screenplays. 
Appendix~\ref{generated-output-script} shows an example of raw generated output.
That said, Dramatron is designed for \emph{interactive co-writing}, as an augmentative tool for human writers.
Co-authorship with Dramatron proceeds as follows:
A writer starts with a log line that they have written. 
They input that log line into Dramatron, and generate a title, characters, a plot outline, location descriptions and each scene's dialogue step-by-step. 
At each step, the writer can take one, or several, of the following operations, as many times as desired:

\begin{itemize}
    \item Generate a new suggestion (i.e., run the LLM again with the same prompt).
    \item Continue generating from the end of the previous generation, similarly to typical ``flat'' LLM generation.
    \item Manually edit some or all of the output generated by the LLM.
\end{itemize}

The writer can furthermore perform these operations by stepping forward and back in the Dramatron hierarchy. For example, they could: 1) generate a title, 2) generate a new title, 3) edit the title, 4) generate a list of characters, 5) edit the characters by removing one character and changing the description of another, 6) generate a plot outline, 7) edit the plot by removing part of the narrative arc, 8) generate a continuation of that edited plot, 9) go back and rewrite the log line, etc. 
This co-writing approach allows the human and Dramatron to both contribute to the authorship of a script. 
Following these operations, the human author could further edit and format to finalize a script.
Appendix \ref{edited-output-script} shows examples of human-edited scripts.

\subsection{Implementation Details}

The code of Dramatron is implemented in Python and the user-facing interface was implemented in a Google Colab\footnote{\url{https://colab.research.google.com/}} with text widgets, allowing interactive editing. 
There are several special markers we use for script generation: \texttt{<end>} represents the end of full sequence generation token, and \texttt{<stop>} is a token used to mark the end of a generated line. 
For a given prompt (see next Sec.~\ref{sec:prompts}) fed to the LLM, up to $511$ text tokens were sampled. 
We used Nucleus sampling~\citep{holtzman2019curious} to encourage diverse outputs, sampling tokens from the top $0.9$ probability mass, and with a softmax temperature of $1.0$. 
Finally, in order to reduce loops in the generation of dialogue, we implemented a simple detector that cuts the generated text into blocks (delimited by $2$ blank lines) and counts how many times each block appears in one single generation. 
Beyond a fixed threshold (e.g., $3$ times), the LLM generates a new output by sampling tokens using a different seed in the random number generator.

\section{Evaluating Text Generated by Large Language Models}
\label{sec:methods}
In this section, we first review existing methods for evaluating LLM-generated text before presenting our approach. 
Echoing~\citet{celikyilmaz2020evaluation}, we split evaluation methods into automated or machine-learned metrics, and human-centric evaluation.
Automated and machine-learned metrics (reviewed in Appendix~\ref{sec:automated-metrics}) typically calculate the similarity between generated and ``ground truth'' stories, the consistency between generated stories and their writing prompts, or the diversity of language in the generated output. These metrics were not designed for generated text of the length of a screenplay or theatre script. This motivates a focus on human-centric evaluation, which can be conducted with \emph{naïve} crowdworkers or with experts.

We now review the limitations of crowdsourced evaluation of the coherence of generated text, and explain why non-expert, crowdsourced evaluation faces crucial quality and bias issues. 
Inheriting from research standards for large-scale natural language processing tasks, the majority of studies assessing the quality of generations from LLMs evaluate model performance by collecting data from crowdworkers  \citep{kozierok2021assessing,nichols2020collaborative,padmakumar2021machine,rashkin2020plotmachines,roemmele2018linguistic,yao2018plan}. For instance \citet{yao2018plan} recruit crowdworkers to evaluate fidelity, coherence, interestingness, and popularity, and \citet{rashkin2020plotmachines} to evaluate narrative flow and ordering of events.
That said, it has been shown that crowdworkers' personal opinions, demographic characteristics and cognitive biases \citep{eickhoff2018cognitive} can affect the quality of crowdsourced annotations in fact-checking \citep{draws2022effects} or in tasks involving subjective assessments \citep{hube2019understanding}.
These issues have led some researchers to try evaluating their models with experts.
\citet{karpinska2021perils} highlight the perils of using crowdsourced workers for evaluating open-ended generated text, because crowdworkers do not read text as carefully as expert teachers. Some studies consulted with expert linguists \citep{du2022read}, university students in the sciences \citep{gero2022sparks} or humanities \cite{yang2022ai}, amateur writers \citep{akoury2020storium,lee2022coauthor,schmitt2021characterchat,chung2022talebrush,yuan2022wordcraft}. In one recent study, \citet{calderwood2020novelists} interviewed $4$ novelists about their usage of GPT-2 via \emph{Talk To Transformer} and \emph{Write With Transformer} tools (see \url{https://app.inferkit.com/demo}), uncovering usages of the ``model as antagonist'' (i.e., random), for ``description creation'', as ``constraint'', or for ``unexpected'' ideas.

Given the issues discussed above, we believe that crowdsourcing are not an effective approach to evaluating screenplays and theatre scripts co-written with language models.
Thus, in a departure from crowd-sourced evaluations, we engage $15$ experts---theatre and film industry professionals---who have both an experience in using AI writing tools and who have worked in TV, film or theatre in one of these capacities: writer, actor, director or producer. 
These experts participate
in a $2$-hour session wherein they co-write a screenplay or theatre script alongside Dramatron. 
Most were able to co-write a full script and the open discussion interview in the allotted $2$-hours; for the others, we slightly extended the interview session.
Interviews are analysed in Section~\ref{sec:results-interviews}, and following the interactive sessions, the participants were asked a series of questions adapted from \citep{yuan2022wordcraft} and detailed in Section~\ref{sec:results-surveys}. 
Each question was answered on a 5-point Likert-type scale, using questions adapted from \citep{yuan2022wordcraft,padmakumar2021machine,clark2018creative,kozierok2021assessing,stevenson2022putting}.
As another quantitative evaluation, we track writer modifications to generated sentences~\citep{roemmele2018linguistic}. This allows for comparison of Dramatron generations pre- and post-human edits.
We track absolute and relative word edit distance, to assess whether and how much the writer adds to or removes from the output suggestions. 
We also track a Jaccard similarity-based metric on words, to quantify how similar is the draft after edits to the original suggestion. 
Our objective is to assess whether Dramatron ``contributes new ideas to writing'', or ``merely expand[s] on [the writer's] existing ideas''~\citep{lee2022coauthor}. We do not evaluate Dramatron outputs for grammatical correctness as~\citep{padmakumar2021machine}, as the few errors made by the LLM can be fixed by the human co-writer.

We compensate experts for the co-writing and interview sessions at $100$ GBP per hour.
Our study design and data collection process was validated and approved by HuBREC (Human Behavioral Research Ethics Committee), which is a research ethics committee run within Deepmind which includes and is chaired by academics from outside the company.

\section{Participant Interviews}
\label{sec:results-interviews}
Throughout our interviews with the $15$ participants (anonymised as p1, p2, etc.), we collected qualitative feedback on co-writing with Dramatron. 
In this section, we summarize this feedback into seven themes. Each is presented alongside supporting quotes from participant interviews. 

\begin{enumerate}
    \item Positive comments about Dramatron focused on: hierarchical generation that lets the writer work on the narrative arc, the possibility either to co-author interactively or to simply let the system generate, and the potential of the output script to serve as source material for the human writer (Section~\ref{sec:interview-positive}).
    \item Participants identified inspiration, world building, and content generation as potential writing applications for Dramatron, and saw it as possible tool for literary analysis (Section~\ref{sec:interviews-potential}).
    \item Participants noticed various biases embedded in the language model (Section~\ref{sec:interview-stereotypes}).
    \item Some writers were interested by the involuntary glitch aesthetic and failure modes of Dramatron, such as repetition and dialogue loops (Section~\ref{sec:interview-glitches}).
    \item Unsurprisingly, participants noticed logical gaps in storytelling, lack of common sense, nuance and subtext, which were manifest in the lack of motivation for the characters (Section~\ref{sec:interviews-limits}).
    \item Structural criticism focused on the need to come up with a log line, as well as on the inconsistencies between consecutive scenes due to parallel dialogue generation (Section~\ref{sec:interviews-structural}).
    \item Participants were engaged with the tool and eager to provide suggestions for improvement (Section~\ref{sec:interview-improve}).
\end{enumerate}

\subsection{Positive Comments about Dramatron}
\label{sec:interview-positive}

\subsubsection{Praise for the interactive hierarchical generation in Dramatron}

All participants but p4 and p5 (who preferred a nonlinear writing workflow) were enthusiastic about the interactive hierarchical generation.
``Once I see this, I know the shape of the series. I know the way that the story unfolds. I can see the narrative more clearly [...] I like this approach of making it a log line and then packing the detail inside it. You are planting a seed of an idea and it is putting meat on the bones'' (p13). 
``All of it is quite consistent, symbolically consistent and coherent and relates to the state of affairs of the state of the play [...] There is lots of emotion and content about relationships in some of the generations'' (p8). 
``In terms of the interactive co-authorship process, I think it is great [...] '' (p9). 
``What I like about the hierarchy is that you can do as much human-ing as you want at any level'' (p2). 
``In working with the machine I can see the content a little more clearly. As there is specificity, character arcs, then I can see how the story comes together [...] This [hierarchical generation] really felt so much cleaner than the process [GPT-2 or GPT-3 with flat prompting] I was using'' (p15).
``Let's try more! God, you could just waste your time doing this'' (p3).
Participants p1, p6 and p3 further noted how such hierarchical generation helped with dialogue: ``there is good content from any generation'' (p1) and (referring to one of the generations) ``You got some big profound discussions in it. I am impressed with that one'' (p3).

\subsubsection{Ease of use of Dramatron's UI and seed-based generation}

Participant p13 liked the user experience of interactive, step-by-step generation of title, characters and plot, whereas p10 thought that ``interaction seemed simpler when the whole script was generated ahead of time rather than editing it''.
Participant p1 tried and discussed three different modes of script generation: 1) interactive co-authorship, 2) modifying the output from one fully automated generation, and 3) curating and modifying outputs from 3-4 generations.
The benefits of running multiple generations included having ``lots of material'', allowing to ``pull good ideas'', ``cherry-picking'', ``more interpretations and artistic freedom'' but ``requires more massaging on my end'' and ``word crafting to make it flow'' (p1). %
Participant p1 developed a workflow for co-generating a script that included editing lists of characters and editing the log line to add more ``characters that we know about'', giving the characters status and names, adding them to the plot's beats. 
When crafting the log line, p1 wanted to imply high stakes and ``stay with humanoid characters: non-human characters take us to the Theatre of the Absurd, to the Surreal, to Magical Realism'', and they wanted log-lines that situated the story in realism ``to meet the audiences expectations'' and ``set things at a specific location''.

\subsubsection{About the potential for the script to be staged after editing}

Several participants (p6, p9, p11, p13, p15) highlighted the potential for the script to be staged after editing: ``a rough draft, would need to work a lot with it [but] it could be helpful and staged, definitely'' (p6), ``It gets me thinking about how you can make a full show with a single idea'' (p11) and ``You know, with a bit of editing, I could take that to Netflix: just need to finesse it a little bit'' (p9). 
Participant p1 staged several scripts generated with Dramatron (see Section~\ref{sec:staging-evaluation}).

\subsection{Potential Uses of the System}
\label{sec:interviews-potential}

\subsubsection{Inspiration for the Writer}
All participants found Dramatron useful for getting inspiration: ``this is perfect for writers' block'' (p13), ``I can see it being very helpful, if you are stuck'' (p4, p5), ``more in depth than the writers' unblocking prompts website'' (p3). 
Dramatron was described as a tool that indirectly stimulates the playwright's creativity: ``I like what happens in my brain when I read some outputs of the model. 
I got an idea for the rest of the story'' (p6), ``It is about me discovering what will translate from what it gives me'' (p10), or that directly gives actionable suggestions: ``Here is a concept; it puts meat on the bones, and then you trim the fat by going back and forth'' (p13). 
Glitches and language model limitations can be subverted for inspiration, in particular when the script is performed: ``mistakes are gifts that we can leave for the improvisers'' (p1).

\subsubsection{Generation of Alternative Choices and World Building}
More than merely providing a creative spark for the main story, the model can be employed to populate the universe of the story: ``If I was going to use this to write a script, I'd use it to generate characters to see if it generated things I hadn't thought about. Or relationships I hadn't thought about'' (p15). 
Dramatron for exploration: ``I would go with the suggestion that is further away from what I would have suggested because I already know what is in my head and I want to know what the machine would do'' (p12).

\subsubsection{Using the System for Learning and Analysis}
By prompting the system, writers could indirectly search the language model for literary styles and elements: ``Even if I were not writing, it does a wonderful job of collecting what is in the literature'' (p10) or even hypothetically search within their own output: ``I would be very interested in feeding everything I ever wrote and then getting it to generate script in my voice and style'' (p4, p5). 
Learning could also happen by analysing how to improve Dramatron's outputs: ``For me, as a playwright, the interesting thing about working with this technology is thinking about how I would edit it. 
For instance: What would this look like on stage?'' (p8).
 
\subsubsection{Content Generation}
Beyond inspiration, several participants were interested by the co-writing potential of Dramatron, and thought it could provide them with material. 
``One of the big sticking points of playwriting is getting words on the page. This helps with that step'' (p8). 
``I would use this tool to fix (screenwriting) projects that might be dead'' (p14).
``This is a rich tool for basically everything. I have done devised creation. There are methods that you can use to generate text, where you pull songs, scripts, or news articles, then chop and paste them down. This reminds me of Dadaist text generation'' (p11). 
``Practically, it might impact the economics of writing if longer running series could be augmented by such writing systems. It might be useful on long-running series, where you have a writers room'' (p4, p5). 
 
\subsubsection{Potential of AI as Tool for TV Screenwriting}
Some participants suggested this tool could be employed in a TV writers' room, to help with writing formulaic scripts.
``If you were able to make an AI to synopsize scripts effectively, you would be valuable to the studio'' (p14).
``It is like having a very good dramaturge'' (p10).
``AI can come up with 5 scripts in 5 minutes'' (p9).
``Which part of the process is this tool relevant for? Formulaic TV series'' (p4, p5).
``I wouldn't use it for writing a straight play'' (p11).

\subsection{Stereotypes}
\label{sec:interview-stereotypes}

\subsubsection{The system outputs are too literal and predictable}

Some participants found the character ``relationships so tight and prescriptive'' (p4, p5); if a character has ``a noble endeavour, it will be stated in the dialogue'' (p4, p5), and that characters were given ``silly'' and ``on the nose, pun names'' (p2). 
Similarly, the title generation ``does what it says on the tin'' (p15), and ``can be overly descriptive sometimes: the director could make decisions'' (p8). One commented, ``this is a thing that my students would do'' (p8).
There were some positive aspects to such a predictable system: ``interpersonal relationships created here are classic tropes that keep the audience interested'' (p3) and ``there is interest in generating outputs from the system for content that already exists: actual titles are fun to compare against'' (p14).

\subsubsection{The system outputs can be problematic, stereotypical, and biased} 

Participant p9 wondered ``What cultures and languages the books come?'' whereas many participants noticed gender biases and ageism in the system outputs. 
``I am less sexist than the computer'' (p3). 
``The protagonists are both male characters, and all of the supporting characters are female'' (p4, p5).
``The female lead is defined by their relationship to the other characters: it is a typical thing in plays that the women characters don't have a lot of information about them'' (p11). ``She is always upset and doesn't have wants (like the male characters) [...]
Actually lots of the content [...] is misogynistic and patriarchal'' (p8).
This problem raised the issue of coping strategies or cultural appropriation: 
``if we gave GPT-2 some character names, it could come up with bigoted characters: [we] went with more made up names, not gender specific, not ethnicity-specific'' (p13) and ``there is an ethical question about using AI for a group of theatre makers: the AI throws us a topic, or relation that is unrelated to our lived experience and we are compelled to \emph{Yes, and} the offers'' (p4, p5). We discuss ethical issues raised in discussion by participants in greater detail in Section~\ref{sec:ethics}.

\subsection{Glitches}
\label{sec:interview-glitches}

\subsubsection{Participants embrace unexpected outputs from the system}

Participant p6 laughed at the ``poetic and absurd'' suggestions. ``It is really interesting to see what it comes up with'' (p8), ``levels of absurdity that are tickling my fancy'' (p10), ``I wouldn't have thought of that but it is quite funny'' (p11). ``This is something that a human author probably would not stand for, it is uniquely created [...] I want ideas that a human couldn't possibly have'' (p12).

\subsubsection{The system often enters in generation loops}

All participants noticed how the system could enter generation loops: ``I would probably cut a lot of it'' (p6) or ``a whole scene about a boiler being broken: yeah'' (p8). They sometimes found positive aspects to such loops: ``It is a silly conversation. It is a little repetitive. I like it.'' (p6), ``repetition leaves room for subtext'' (p12) and enjoyed the glitches (p4, p5) or even made parallels with existing work (p3).

\subsection{Fundamental Limitations of the Language Model and of Dramatron}
\label{sec:interviews-limits}

\subsubsection{Lack of consistency and of long-term coherence}
``Keeping dialogue character-based and consistent is most important [...] There is still some difficulty in getting it to stay on track with the context.'' (p15).
``I want the characters to be more consistent within themselves'' (p12).
``There is a bit of confusion in the logic, gaps in logic [...] It looks like postmodern theatre [...] But in terms of [a play with a given] genre, that has a plot to follow, it is getting confusing'' (p11).
Participant 7 ``wants to add some stitching between the beats to make them narratively make sense''.
 
\subsubsection{Lack of common sense and embodiment}
Participant 8 observed that ``There are things that it is hard to show on stage – such as a cat. The system doesn't have an awareness of what is stageable and not stageable'' and p9 noted that when ``interfacing with a story telling AI, the input space is constrained''.

\subsubsection{Lack of nuance and subtext}
Participant 3 observed: ``that's a good example of how computers do not understand nuance, the way we see language and can understand it even if it is not super specific''. ``A lot of information, a bit too verbalised, there should be more subtext'' (p6).
``With dialogue in plays, you have to ask yourself two questions: 1) Do people actually speak like that? 2) Are actors attracted to these lines and are these appealing lines to play?'' (p7)
``Playwriting is about realistic dialogue... all of the things around subtext. [...] Show, not tell: here we are just telling. Just like in improv: `do not mention the thing'. The element in the log line became the central bit in the generation, and that was repetitive'' (p8).
Participant 14 concluded that
``AI will never write Casablanca, or A Wonderful Life. It might be able to write genre boxed storytelling''.

\subsubsection{Lack of a motivation for the characters}
``The stories do not finish. The character journeys are not complete. There is perhaps something missing in the character background [...] Where is the emotional motivation, stuff that might exist in the backstory and not exist in the script?'' (p14).
``On the first go-through, you are looking for the goal of the protagonist, and impediment for that drive.  What is my character doing, and what do they want? If this was given to an actor they are going to struggle with the first thing to do, which is to find the needs and the wants of the character and then to personalise it'' (p9).
``My students do this: a character comes into play and says right what they want.'' (p8).
``The conflict should be something inside the character'' (p6).
``Why do people not say what they mean? It is because we have societal understanding, but sometimes get lost in translation'' (p3).

\subsection{Structural Problems of Dramatron}
\label{sec:interviews-structural}

\subsubsection{Difficulty caused by the need to come up with the log line to condition all the generation}
For participant 12, it was difficult to come up with a log line, and the process seemed precious. ``Coming up with the first prompt takes a little bit of back and forth'' (p11).
``Packing the action into the log line: this is a panic moment for the writer, because they want to add everything meaningful into the script. [...] It is all about the witty premise. The system that you have right now is somewhat about wit. There is a need for the log line to hold some kind of wit'' (p13).
``Does [the log line] have to have a character name? (p4, p5).
``The log line is not a closed synopsis. It is less descriptive and more prescriptive. The art of log lines is about how short you can make it so that [the producers] read the rest of your material'' (p14).

\subsubsection{Structural criticism of log line-based conditioning of the whole generation}
``Generally the way that I work, I am clear what I want to say about the world – what I think about the world. The vehicles, or the characters, or the arc is not clear. This looks like a collection of scenes that logically follow one to the next. But, the core idea of the thing to say [is missing]'' (p4, p5).
``If I could program something to write a script, I wouldn't start with a log line. You can also consider starting with a character and an obstacle in the way of that character'' (p9).
 
\subsubsection{Negative consequence of Dramatron's design choice: parallel dialogue generation}
``From the scene beats, it has no idea of what the previous dialogue contained. Then to see the dialogue not be consistent is jarring'' (p1).
``I wonder if there is a problem in importing the previous beat into the scene [...] Paying attention to the consistency in the beats, helps with the consistency of the dialogue generated'' (p12).
Upon learning that scene dialogue was generated in parallel for each scene, Participant 2 commented: ``If it didn't read its last scene, how can you get the last scene into the next generation? Generation of these scripts could be significantly benefited from attending to the previous scene's dialogue''.

\subsection{Suggested Improvements to Dramatron}
\label{sec:interview-improve}

Modeling characters and their relationships was a recurrent theme: ``can we make the system relationship-driven?'' (p12), ``where does status belong in character building?'' (p12), ``could we generate the stem of a character and then complete it?'' (p15). Participant 12 suggested: ``as an author, I would build a social graph of the characters relations''.
Answering the question ``How do you get the system to know where the scene should start and end?'' (p15), three participants (p8, p13, p15) suggested fitting a narrative arc within each scene.

Several participants wanted to be able to query and dialogue with the writing model: ``Have you engaged [the AI system] by trying to give it notes?'' (p2) to allow it to learn about the world: ``How does world building happen? Maybe the model needs to know the Ws of Stella Adler [(Who? What? Where? Why? How? etc.)] Can you get the system to answer these questions?'' (p9), or to allow rewriting and reformulation: ``can we ask the system to re-write with a style or context?'' (p8). 
As p10 reiterates, iterative rewriting was a desired workflow: ``I am less interested in shaping [the narrative], rather than seeing what it is saying, and refining it to see what it says, and then refining it again. A playwright has to see the play spoken before making cuts.''

Finally, p4 and p5 astutely observed that ``there has been a push away from systems of Western dramaturgy, so in terms of making this most useful for the future, it might be helpful to consider how it might be used within the context of other contemporary writing''---suggesting alternative narrative structures and elements---``as the AI is not bound by the same rules that we are. So, telling it to be bound by those human rules feels limiting of the capabilities''.

\subsection{Incremental Tool Improvement}

As detailed in Section~\ref{sec:interview-improve}, the participants were engaged and provided constructive feedback about Dramatron. 
As one of the participants in the study remarked: ``the system is so adaptable, it can change with our feedback and tweaks''. 
This sort of understanding of the systems modifiability empowered those that interacted with it to more freely suggest changes, knowing that they could be incorporated. 
In this way, the system positively benefited and evolved over the course of the participant study. 

Over the course of the interviews, we incorporate the feedback we could by making small, incremental changes to the prompt prefix sets of Dramatron. 
Table \ref{tab:changes} summarizes changes made as a direct result of participant's feedback.
This sort of participatory design and development is critical for creative tool generation as the feedback from users can be directly incorporated to improve the system for the next interaction.
This is made possible via the modular design of the system, the lightweight prompt-based interactions, and the flexibility afforded by Dramatron. 
This participation also inspires participants to explore related, connected, creative ideas. 
For example Fig. \ref{fig:production_images} (LEFT) shows concept art for a narrative test of virtual actors interpreting a co-written script.

\begin{table}[ht!]
\begin{tabular}{lll}
Participants & Prompt set & Modifications of Dramatron subsequent to the sessions \\ \hline \hline
p1 (session 1) & Medea & 
Pause after each scene dialogue generation to enable review by the writer. \\
\hline
p1 (session 2) & Medea & Simplify plot prefixes to location, narrative element, beat, list of characters.  \\
 &  & Use an edited script generated by Dramatron (``The Cave'') as prefixes. \\
\hline
p1 (session 3) & The Cave & --- \\
\hline
p2 & The Cave & Continue the generation of dialogue. \\
 &  & Enable editing of generated dialogue. \\
 &  & Rewrite the title prefix to avoid plagiarism. \\
\hline
p3, p4, p5, p6 & Medea & Make scene dialogue prefix depend on current and previous scenes' beat summaries. \\
 &  & Continue the generation of characters and plot. \\
\hline
p7, p8 & Medea & Remove the list of characters from plot prefixes. \\
 &  & Show multiple seeds for title outputs. \\
 &  & Sample new generation in case a loop is detected in generated dialogue. \\
\hline
p9 & Medea & Added Sci-Fi prefixes  \\
\hline
p10, p11 & Sci-Fi & --- \\
\hline
p12-p15 & Medea & --- \\
 \end{tabular}
\caption{Summary of prompt sets used and incremental tool improvements following participant sessions.}
\label{tab:changes}
\end{table}

\subsection{Staging and Evaluating Productions of Scripts Co-written by Dramatron}
\label{sec:staging-evaluation}

Creative writing for theatre is fundamentally interactive: not just between collaborating storytellers, but between storytellers and the audience.
For this reason, we evaluated how scripts co-written with Dramatron could be produced on the theatre stage. 
In this section, we describe staging details and report evaluative reflections from both the creative team and two professional theatre reviewers.

\begin{figure}[!t]
\includegraphics[width=0.395\linewidth]{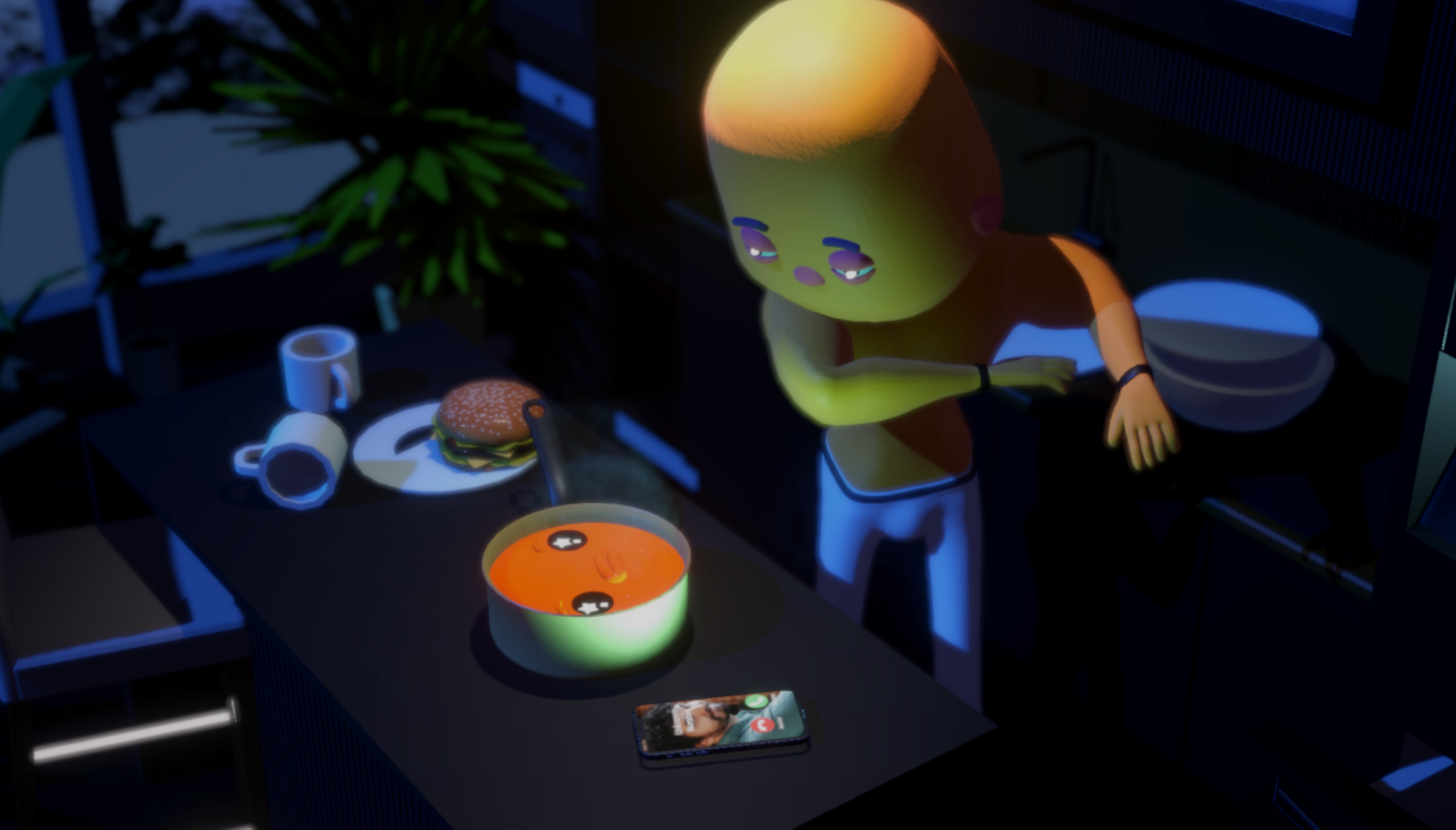}
\includegraphics[width=0.5\linewidth]{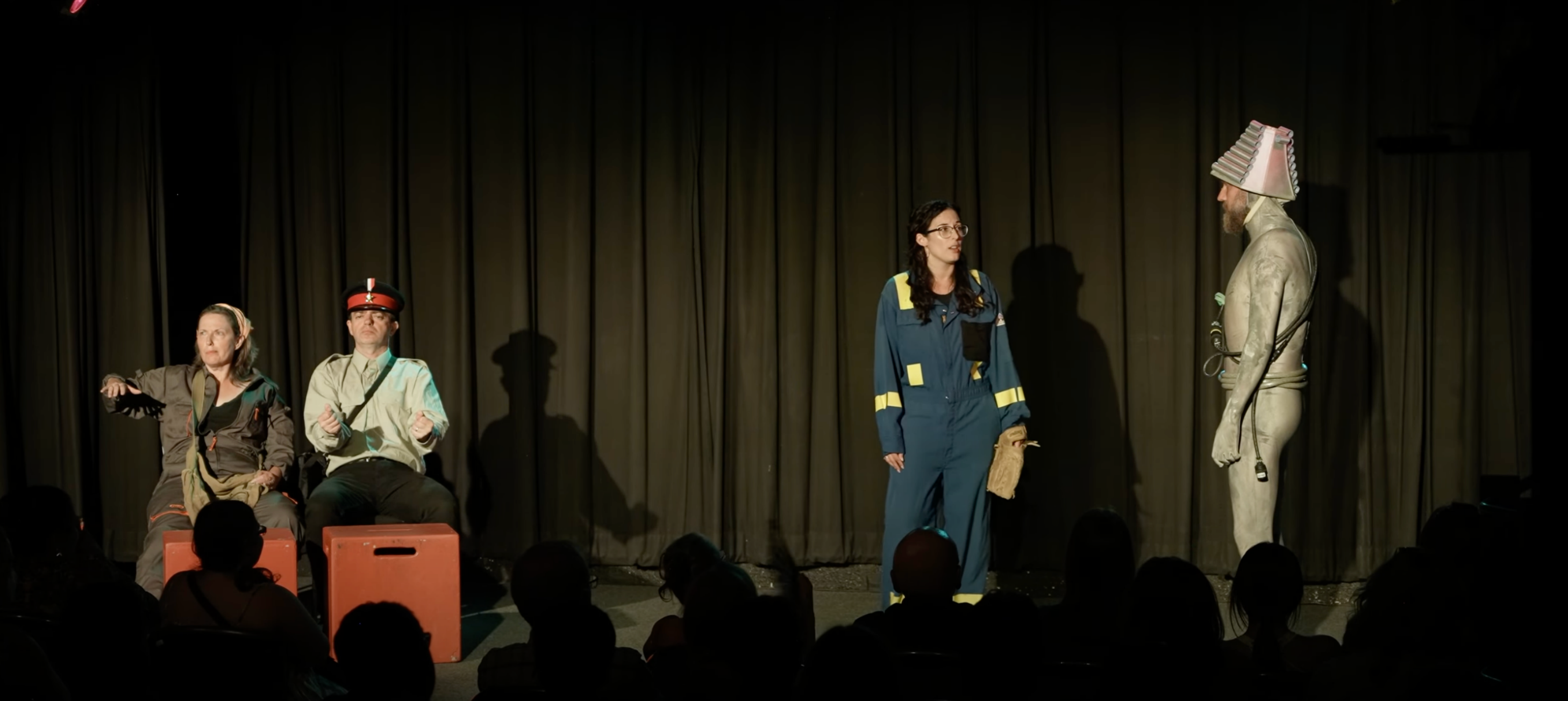}
\caption{
    (LEFT): Concept art used for a narrative test prototype of a virtual actor interpretation of the script \textit{Darren just can't handle the temperature of his soup} created by Participant p13. Used with Permission from Transitional Forms.
    (RIGHT): Photo of human actors interpreting the script~\textit{Cars: The Day The Earth Stood Still} as part of the \emph{Plays By Bots} series of performances of scripts co-written with Dramatron and director Participant p1. Used with Permission from Rapid Fire Theatre.
}
\label{fig:production_images}
\end{figure}

Five scripts co-written with Dramatron were staged in public performances in August 2022 at North America's largest theatre festival: The 2022 Edmonton International Fringe Theatre festival.
The show's run was titled \textit{Plays By Bots} and ran $7$ performances over two weeks (see an image from the production on Fig.~\ref{fig:production_images}). 
In each show, different casts would act out one of the plays from the co-writing experiments. 
The plays span different genres, styles, characters, and storylines. 
The scripts were brought to life by a cast of 4-6 experienced improvisers and actors.
The first-half of each script was given to each of the cast members in a sealed envelope. 
Only when the show began were they allowed to open the script, and then they commenced performance by reading it live in front of the audience. 
Once the script ran out, the actors improvised the ending, based on the context and story set out by the script\footnote{Video of performance shared upon acceptance.}.
During each show's performance, the director and co-writer (participant p1 from above) introduced the project to the audience and explained that they co-wrote and edited the script using Dramatron.

There were two reviews written about the production of \textit{Plays By Bots} at the festival. One of the reviews noted that the performance ``proves that artificial intelligence can in fact write a hit Fringe play''. 
The reviewer also noted that the success of the performance was due to both the Dramatron system and the human actors, especially one performer who ``mastered Dramatron's voice and seamlessly took it off-script for the remainder of the show, much to the delight of the howling audience''.
The second review was also positive. With a hint of incredulity, the reviewer complimented the abilities of Dramatron.
The reviewer noted the style of Dramatron, and how that served the performance saying ``if there's a certain flatness in the dialogue, which runs to declarations, that in itself is amusing since it turned out to be perfectly suited to the deadpan comic talents of [the] improvisers,'' and ``the human actors continue to capture the playwright bot's tone''.
The reviewer also expressed surprise at the ability of the system to create a play that hangs together and creates a world. 
They further noted that some lines from Dramatron are so funny they were reprised later in the show once the human actors were improvising.

Discussions amongst the creative team compliment the reviewers and provide insights on how professional actors and improvisers found working with scripts co-written by Dramatron. 
Post-show discussions were facilitated and relayed to us by the director (p1 above). 
Four key themes emerged through these discussions which echo the themes presented earlier in Section ~\ref{sec:results-interviews}. Specifically, the system has a distinct glitch style, generated text can be repetitive and fun to work with. 
As well, the team attributed agency to the system, and had expectations of the systems capabilities. 
As trained improvisational theatre performers, the actors were able to add a layer of interpretation to the co-written script. 
This helped add meaning to the text.
Finally, the prevailing feedback from the creative team was that participating in the production was fun! 
Enthusiasm and reflections from the creative team echo the usefulness of co-written scripts for theatre production and collaboration; more reflections and supporting quotes are included in Appendix \ref{sec:appdx-plays-by-bots}.

\section{Participant Surveys}
\label{sec:results-surveys}
\subsection{Qualitative and Quantitative Results on Participant Surveys}

Of the total $15$ study participants, $13$ provided responses on our post-session feedback form. The form gave participants the following instruction: ``When answering these questions, please reflect on the interactive co-authorship session as well as considering the use of an interactive AI system like Dramatron in the future'', and asked nine questions. Each question could be answered using a Likert-type scale ranging from Strongly Disagree $1$ to $5$ Strongly Agree. These questions are slightly adapted from~\citet{yuan2022wordcraft} and ~\citet{stevenson2022putting}:
(1) \emph{I found the AI system helpful},
(2) \emph{I felt like I was collaborating with the AI system},
(3) \emph{I found it easy to write with the AI system},
(4) \emph{I enjoyed writing with the AI system},
(5) \emph{I was able to express my creative goals while writing with the AI system},
(6) \emph{The script(s) written with the AI system feel unique},
(7) \emph{I feel I have ownership over the created script(s)},
(8) \emph{I was surprised by the responses from the AI system}, and
(9) \emph{I'm proud of the final outputs.}
We also asked five free-form questions. Two questions aimed at assessing the participants' exposure to AI writing tools (\emph{In a few words: what is your experience in using AI tools for writing for theatre of film or during performance on stage?}) and their industry experience (\emph{In a few words: what is your experience in theatre or film/TV?}). We used these questions to manually define a binary indicator variable (\emph{Has experience of AI writing tools}) and a three-class category for their primary domain of expertise (\emph{Improvisation}, \emph{Scripted Theatre} and \emph{Film/TV}). Three more questions gave participants an opportunity to provide developmental feedback about the system: \emph{What is one thing that the AI system did well?}, \emph{What is one improvement for the AI system?} and \emph{Please provide any comments, reflections, or open questions that came up for you during the co-authorship session or when answering this survey}.

\begin{figure}
    \centering
    \includegraphics[width=0.4\textwidth]{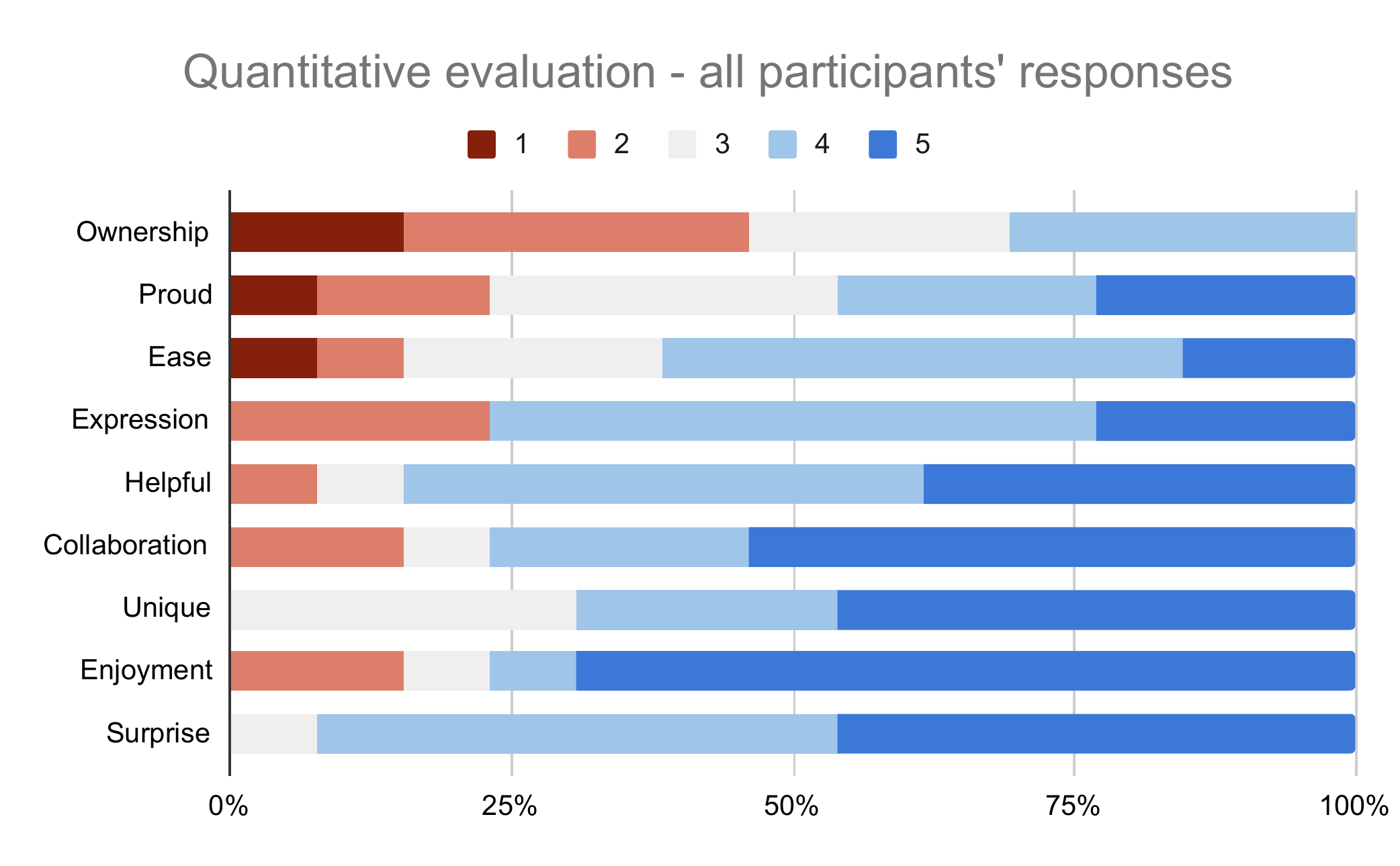}
    \includegraphics[width=0.4\textwidth]{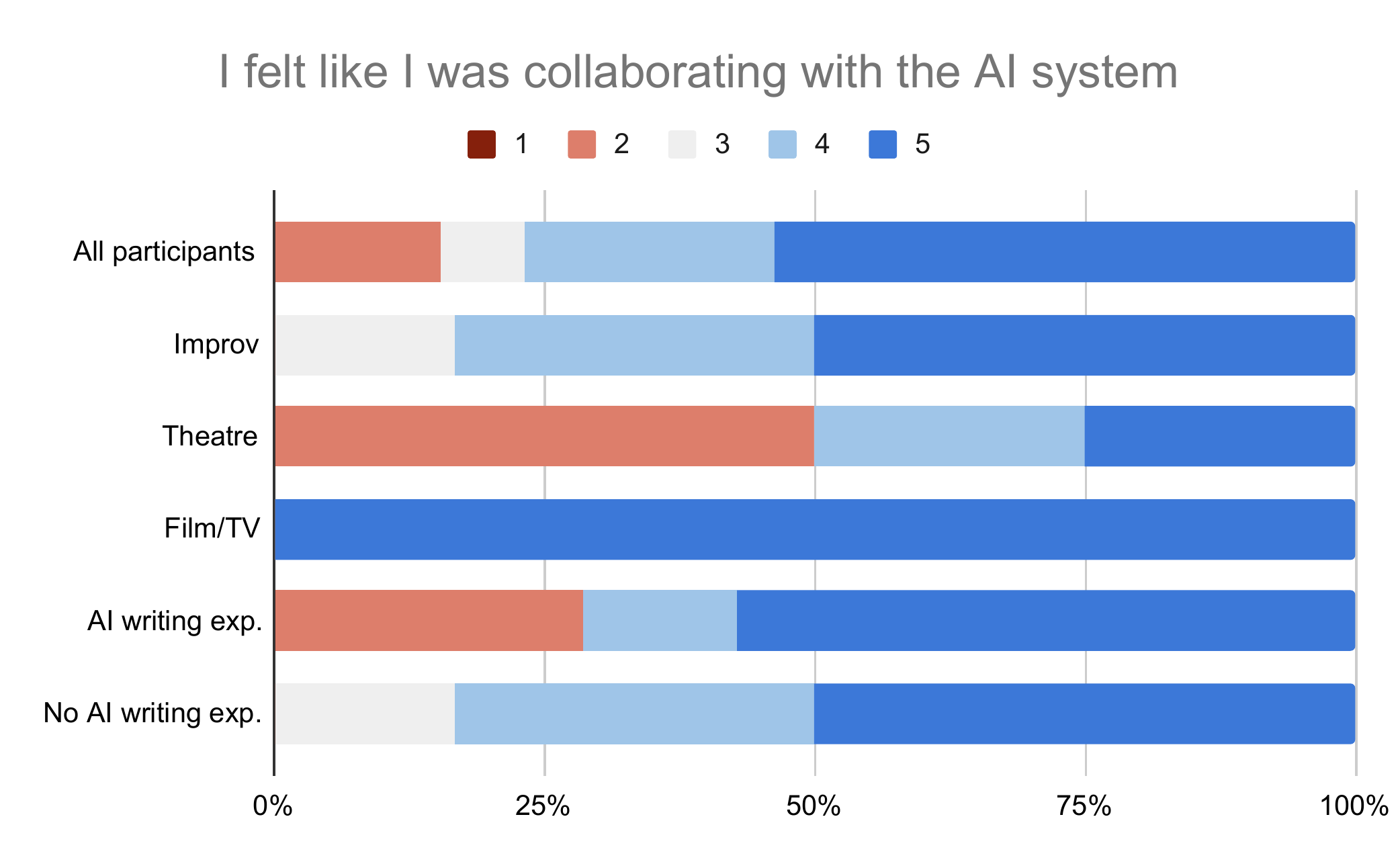}
    \caption{Participants responses to the quantitative evaluation, on a Likert scale from 1 (strongly disagree) to 5 (strongly agree)}
    \label{fig:likert}
\end{figure}

Aggregated responses are plotted on Figure \ref{fig:likert} (Left), and (Right) factors responses to the question on \textbf{collaboration} by industry background and experience (see Figure \ref{fig:likert-supp} for more). The results are summarized below.

\subsubsection{AI-generated scripts can be surprising, enjoyable, and unique}

The statements that received the strongest positive response were \emph{I was {\bf surprised} by the responses from the AI system} (92\% participants agreed, 46\% strongly agreed) followed by \emph{I {\bf enjoyed} writing with the AI system} (77\% agreed, 69\% strongly agreed) and \emph{The script(s) written with the AI system feel {\bf unique}} (69\% agreed, 46\% strongly agreed), referring to lines of dialogue or connections between characters.

\subsubsection{Dramatron can be helpful for expressing creative goals for screenwriting}

Positive responses to statements \emph{I felt like I was {\bf collaborating} with the AI system}, (77\% agreed, 54\% strongly agreed), \emph{I found the AI system {\bf helpful}} (84\% agreed, 38\% strongly agreed) and \emph{I was able to {\bf express} my creative goals with the AI system} (77\% agreed, 23\% strongly agreed) suggest a fruitful interaction between a writer and an interactive AI writing tool, in particular for ideation, through multiple choices, twists to the narrative, or specific details.

\subsubsection{Dramatron outputs were preferred for screenplays over theatre scripts}

We noticed that the subgroup of respondents with a film or TV background gave the highest scores for the surprise, enjoyment, collaboration, helpful, expression, ease and ownership questions. 
Respondents with theatre background judged enjoyment, collaboration, helpful, creative, expression, ease, proud and ownership the least.
This reinforces our observations during interviews (see the criticism of log line-based conditioning for generating theatre scripts in Section~\ref{sec:interviews-structural}). %

\subsubsection{Dramatron's hierarchical prompting interface is moderately easy to use}

Participants were more split about the ease of use of the tool (61\% agreed) and whether they felt proud about the output of the system (46\% agreed), with one participant strongly disagreeing. We note that these two answers were highly correlated ($r=0.9$), and that several questions had a relatively strong correlation ($r\geq0.66$): helpful, collaboration, ease, enjoyment, expression and ease. As observed during interviews, recurrent criticism was about Dramatron needing detailed log lines to start the generation.

\subsubsection{Participants felt a low level of ownership for the co-written scripts}

More than half of respondents felt they did not own the result of the AI system. One participant commented that the generated script was only a starting point provided by the AI and that the writer still needed to create their own script.
One interpretation is that Dramatron is seen by the industry experts not as much as a full script writing tool but rather as 
a learning tool to practice writing and a source of inspiration that generates ``\emph{not necessarily completed works so much as provocations}'' (p7).

\subsection{Quantitative Observations}

We observed that participants would skim the text and remark on specific inspiring lines, but were less focused on the overall \emph{coherence} of the discourse. 
That said, participants did notice if the dialogue of a scene was \textit{unrelated} to that scene's beat or log line, and most noticed when the generation had excessive \textit{repetition}.
Thus, we pair our qualitative results with a small set of descriptive statistics measuring these features and relevant to our participants' feedback.

Lemma-based Jaccard similarity is a metric measuring the overlap between two sets of lemmatised vocabularies.
We found that when participants made multiple generations for a scene description, they generally did not choose outputs with greatest Jaccard similarity.
We performed a one-sided Wilcox signed-rank test on Jaccard similarity score differences, testing whether chosen output seeds were more similar to the log line provided by participants.
We found that the beats for chosen generations are not more similar to the log line when compared to the beats for generations not chosen (\textit{W} = $28$, \textit{p} = $0.926$). 
In fact, there is a strong trend in the opposite direction. 
This suggests that word overlap is not predictive of which output generation is selected.

\begin{figure}[]
\centering
\includegraphics[width=0.33\textwidth]{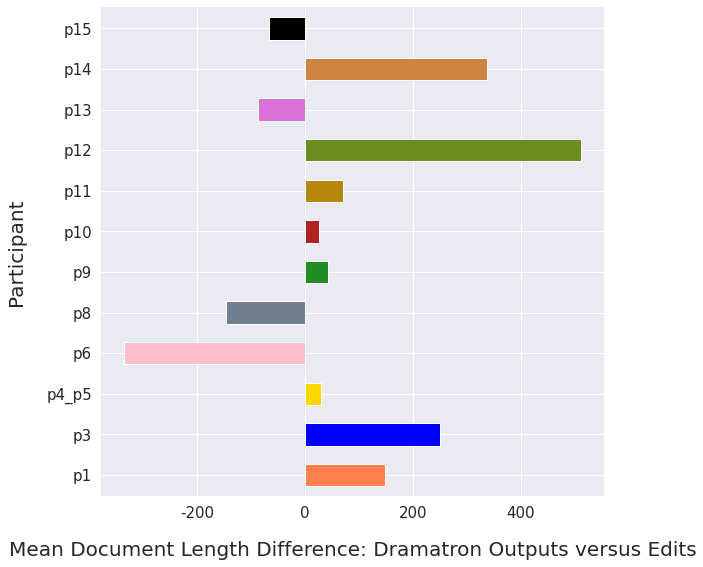}
\includegraphics[width=0.31\textwidth]{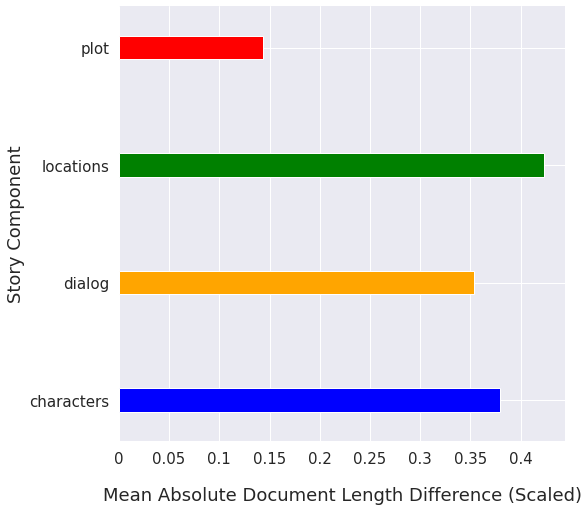}
\includegraphics[width=0.283\textwidth]{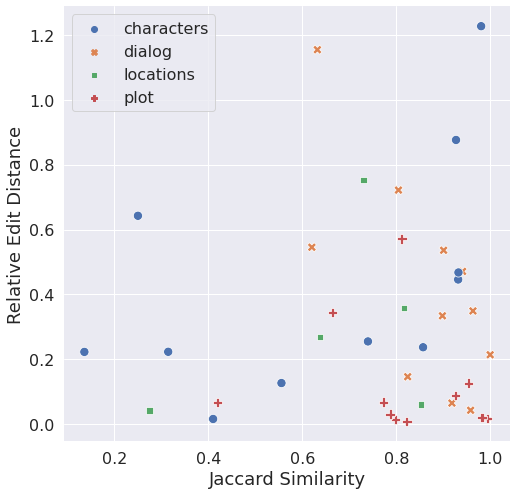}
\setlength{\belowcaptionskip}{-20pt}
\caption{
Left: Mean length difference between Dramatron output and final edit, by participant.
Middle: Normalized mean absolute length difference by story component. 
Right: Scatter plot of relative edit distance and Jaccard similarity for edited texts versus original Dramatron outputs.
}
\label{fig:edits}
\end{figure}

Using an open-source repetition scorer~\citep{pegasus},
we examined the repetition scores for texts where participants generated multiple dialogue outputs but chose only one to incorporate into the script ($n=3$).
After computing a one-sided Wilcox signed-rank test, we found no significant difference between the repetition scores for chosen Dramatron output against alternative seed outputs (\textit{W} = 57, \textit{p} = 0.467). 
This finding aligns with our observations in the interactive sessions where participants tended to delete degenerate repetitions, or as one participant (p13) put it: "[Dramatron] puts meat on the bones... And then you trim the fat by going back and forth."
We interpret our distance measures as indicators of engagement with Dramatron. 
Figure~\ref{fig:edits} (right) displays mean relative Levenshtein distances for each participant, along with Jaccard similarity.

We also report document length differences, to show the directionality of editing (i.e. whether participants are adding or deleting material to or from Dramatron output.
Figure~\ref{fig:edits} (left) shows the mean document length difference by participant. 
Figure~\ref{fig:edits} (middle) shows these measures normalised and grouped by type of generation. 
We notice that four participants tended to add to generated outputs (three of them, p6, p13, and p15 expressed interest in staging the resulting scripts) 
while eight others tended to remove from and curate the outputs (including p1 who staged productions).
We also notice that the plot outline was the least edited part of the script. 
During co-authorship we observed that participants tended to delete longer location descriptions and parts of dialogue (especially loops), and rewrite the character descriptions.

\section{Discussion and Future Work}
\label{sec:disc-and-conclusion}

\subsection{Future Work Towards Coherent Story Generation}

While common sense and logical consistency is an elusive goal for LLMs~\citep{bender2021dangers}, their utility as a writing tool increases as they generate more coherent outputs.
Hierarchical generation can be leveraged in various ways to improve the long-range coherence of long texts.
One enhancement to improve coherence, especially suited for screenplays and theatre scripts, is a method to generate complex and complete character arcs.
Likewise, generating satisfying scene conclusions in dialogue is rare. 
One technique to address this is to use hierarchical dialogue generation by constructing each scene's dialogue from a generated beginning, middle, and end dialogue beat.
Finally, to improve stylistic coherence, future work could explore methods to generate thematic outputs satisfying notions of genre. 
This could be done by writing new prompts or transposing existing ones into a variety of diverse author styles and voices.

\subsection{On the Difference between Film and Theatre, and a Critique of the Formulaic Hierarchy in Dramatron}
\label{sec:difference-film-theatre}

As Dramatron relies on top-down hierarchical generation, each subsequent step depends on those that came prior. 
Several participants noted that this style of writing is more aligned with screenwriting than playwriting.
In the words of some industry professionals, they were ``drawing a line between screenwriting and script writing for stage. Playwrights do not use the log line in the same way as in screen writing'' (p9). 
Participants 4 and 5, who self-identified as playwrights for the stage, ``would never approach a piece of work with a story in [their] head. [They] might come with a more investigative approach with a theme''. 
In fact, they argued against Dramatron's constrained top-down hierarchy: ``Why generate characters first? At earlier parts of the creation process you might not know the characters. In the way [we] work, [we] often come with characters at the end, and the idea of characters comes last''. 
The log line itself could be seen as a post-hoc summary, as ``often times playwrights find out what the play is about after [they] finish it. I will know the play once it is done'' (p9). 
This said, Dramatron does allow for going back-and-forth between levels in the hierarchy, and through creative prompting, future work could allow for generation steps can happen out of order.

Differences in the writing of screenplays and theatre scripts can also be related to cultural and economic factors: 
``The difference between theatre and screen is that nobody's making theatre on demand, no outside pressure. 
Or at least not in the same way that there is for television. Theatre feels less like generating content'' (p4, p5) whereas ``film scripts, in the industry, want the traditional fourth wall'' (p9). 
This reflection invites us to reconsider the applicability of Dramatron. 
Since our tool is formulaic by construction, is it suitable to TV or film production? 
As one respondent noted, ``[Dramatron] will be very useful for an average Hollywood movie and for television. It does not need to have a deep understanding of the human soul, unlike Shakespeare. [...] The thing with action movies is that is that actors are not expected to connect with the writer. A screenwriter on a TV set is just like [Dramatron] [...]. It is a sublime skill to be a Hollywood writer because your creative input is small.'' (p9).

\subsection{Ethical Questions and Risks Posed by AI Writing Tools}
\label{sec:ethics}

We describe a co-creative tool built around large language models. It can augment and uplift human artists' work by providing them with inspiration, as well as challenge them and thereby support their artistic practice.
Before conducting our study, we identified three directly relevant risks and ethical implications discussed in previous work~\citep{weidinger2021ethical}: 
1) bias and offensive language in the generated output, 
2) automation of creative work resulting in ``cannibalizing'' the work of creative artists engaged in script writing, and 
3) copyright infringement by reusing copyrighted data from the training dataset, either knowingly (e.g. through prompting: ``write the script in the style of Ursula Le Guin'') or unknowingly (e.g. by virtue of similar training data). 
Our mitigation strategy is two-fold: we invite the creative human artist into the loop throughout the co-authorship process, and we maintain clarity and transparency on the origin of the generated text. 

To mitigate copyright issues, the writer could query short parts of the script using a search engine and plagiarism detection tools~\citep{lee2022language}; this functionality could be built directly into co-creative tools.
Writers using these tools should be aware of the origin of the data in the LLM, and their audiences should be aware that those outputs were generated through an interaction between humans and co-creative tools. Interestingly, study participants independently raised these concerns during interviews.

From the feedback gathered in the study, some participants reported that outputs from the LLM can sometimes be problematic, stereotypical, or biased: 
for example, ``I am less sexist than the computer'' (p3), or ``the protagonists are both male characters, and all of the supporting characters are female'' (p4, p5).
Furthermore, participants raised concerns about the source of the dataset: 
``If you are putting existing scripts into the dataset, where are they being pulled from?'' (p4, p5).
Thoughts on this subject ranged from ``Plagiarising the corpus of scripts is a problem'' (p2) to ``In the context of collective and devised creation, [reusing existing published work] is not necessarily a problem, because it can be perceived as an homage to existing work'' (p11).
The rules and norms for the use of systems trained on copyright-protected material are the subject of ongoing work~\citep{bommasani2021opportunities}.
For example, Lee \textit{et al.} (2022) distinguish between verbatim, paraphrase, and idea plagiarism~\citep{lee2022language}.
Finally, participants raised concern about the potential impact of generative tools on creative economies:
``It would free the artist from writing formulaic scripts, [but] it also replaces the work opportunities'' (p4, p5).
In general, participants found our mitigation strategies satisfactory and none reported distress or concern regarding outputs from the model. 
While not the prime focus of the interview sessions, biases and stereotypes could be systematically explored: future work could explore what sorts of narratives can be written using using AI tools, and how the system performs for different cultural groups.

\subsection{Using a Tool or Participating in a Co-Creative System?}

Previous work has argued for 
engagement with subject matter experts, literary scholars, and industry professionals~\citep{van2021ai}. 
In this work, screenwriters and playwrights co-wrote with Dramatron. 
In the post-interview surveys, most of the participants felt they did not own the final output.
This raises several questions: 
Should Dramatron be considered merely a writing tool, or should it rather be seen as a co-creative system? 
Are writers comfortable using and ready to adopt co-creative tools? 
Participants reflected on the authorship of machine co-created text (``Interesting what this means for the future of writing. Who is the author?'', p6). 
As a corollary to the issues of authorship and of biases, p3 wondered whether an LLM should generate text from the perspectives of different identities, and if that could amount to cultural appropriation (although they later added: ``but I write about many people too, and I am less objective than this AI, because I have seen less data'').
\citet{chung2022artist} discuss how AI-based Creativity Support Tools can be seen as part of the Artist Support Network. 
These tools may need to conform to the artists' expectations of collaboration, similarly to the types of interactions they have with human collaborators: for example, sub-contracting, co-creation, or inspiration. 
Our expert interviews and surveys surfaced similar views towards the interaction with Dramatron and co-creative writing tools in general.

\section{Conclusions}

We present Dramatron: an interactive co-writing tool which allows writers to generate scripts from a provided log line. 
Hierarchical story generation with explicit narrative structures and characters helps to generate more coherent text, especially when generating text as long as theatre scripts and screenplays.
We conducted a user study with $15$ theatre and film industry professionals and distilled their reflections collected through open-ended qualitative interviews and a short survey.
We also present feedback from a creative team that produced scripts co-written with Dramatron in public performances at a theatre festival, alongside two reviews from professional reviewers.
In summary, Dramatron can be used as a co-creative writing tool allowing human authors to write screenplays and theatre scripts alongside LLMs.  
This work invites further questions on the nature of co-creativity and on the ethics surrounding LLMs.

\clearpage

\section*{Acknowledgements}

We would also like to thank anonymous reviewers for their time, energy, and insightful feedback, as well as our colleagues at DeepMind for creative inspiration and critical input on the scientific, ethical and legal aspects of this work, in particular: Tara Thomas, Kevin McKee, Boxi Wu, Antonia Paterson, Murray Shanahan, Robert Dickens, Aliya Ahmad, Danielle Breen, Sanah Choudhry, Joel Moss, Yan Lai, Jon Small, Will Hawkins, Laura Weidinger, Lisa Anne Hendricks, Mia Glaese, Geoffrey Irving, Jack Rae, Natalie Lambert, Raia Hadsell, Shakir Mohamed and Doina Precup.

We are immensely grateful to the anonymous participants who took part in this study and who made it possible. Finally, we are indebted to the talented performers and production companies Rapid Fire Theatre in Edmonton, Canada and Transitional Forms in Toronto, Canada without whom we would never have been able to fully realise the generated scripts. Thank you for providing your artistic voices in this human-machine co-creative dialogue.

\bibliographystyle{abbrvnat}
\bibliography{dramatron}

\begin{appendix}

\clearpage

\section{Related Work on Automated Story Generation and Controllable Story Generation}
\label{sec:related}

In this section we provide background and related work on the intersecting fields of automatic plot and story generation as well as controllable language generation.

\subsection{Automatic Story Generation}

Automatic story generation is the research problem concerned with generating sequences of story elements that collectively tell a coherent narrative. 
A narrative plot is a sequence of events where each affects the next. 
The plot is composed of narrative elements sometimes referred to as actions, beats, scenes, or events~\citep{meehan1976metanovel}. 
Generative plot systems have been developed for nearly a century by~\citet{cook1928plotto}, and computerized versions have existed for decades~\citep{meehan1977tale}. 
These systems support human authors with creative output material and as a source of randomness. 
Recent work has adapted these systems for computational interaction for use in web-based and theatrical settings~\citep{eger2015plotter,eger2018dairector}. 
In generating narratives, the combinations of these component elements form subplots. 
Multiple subplots can be combined into a single plot, and multiple plots can intertwine to create complex narratives. 
Many contemporary stories are written to have multiple plot lines which intertwine.
But, there is little work on how to computationally model and generate multiple intersecting plot lines.
Complex plot line interaction is a promising avenue of future work for human-machine co-creativity research in story generation. 

Early approaches to automatic story generation used symbolic planning and hand-engineered heuristics~\citep{riedl2010narrative,li2013story,ware2011cpocl,meehan1977tale,theune2003virtual}. 
Recently, research has explored open-story generation using machine learning techniques which leverage large datasets, massive deep learning models, increased compute capacity, and large language model prompt engineering~\citep{fan2018hierarchical,see2019massively,branwen2020gpt,Polceanu2021NarrativePG,reif2021recipe,calderwood2022spinning}. 
These methods show how models can succeed, and fail, in the generation of unique and coherent stories. 
Additionally, while coherence has been studied in dialogue generation methods~\citep{dziri2019evaluating}, it remains challenging to measure coherence in story, specifically as it relates to causal narrative events, or common sense knowledge~\citep{alabdulkarim2021automatic}, or consistency in characters~\citep{papalampidi2022towards}.

\subsection{Symbolic and Hierarchical Story Generation}

Some work has tried to bridge the gap between symbolic event representations and textual representations.
Several of these methods process and predict events from text~\citep{martin2017event,wang2019t,guan2019story,sakaguchi2021proscript} by generating sequences of plot events and then expanding such plot events into sentences~\citep{ammanabrolu2020story,rashkin2020plotmachines}.
Others model each story as a series of character and story challenge cards~\citep{akoury2020storium} (first simulating sequences of causal events, then transforming them into sentences) or by simulating social practices between autonomous agents~\citep{evans2013versu}.

Other recent work separates storytelling into two phases: storyline (i.e. plot) planning and story writing based on that storyline~\citep{yao2018plan}. Similarly, methods have been introduced which decompose story generation into processes of coarse-to-fine generation~\citep{fan2019strategies,celikyilmaz2020evaluation}.
\citet{goldfarb2020content} further introduced rescoring methods for character and plot events. 
These methods have not focused on synthesising coherent stories by generating scenes and dialogue, as we do in our work. 
Many of these methods lean on human reading comprehension and preference-based evaluation as opposed to production of the final script.

Hierarchical generation of a theatre play was first mentioned and used in \citep{rosa2022gpt,rosa2020theaitre} for the production of \emph{AI: Can a Robot Write a Play?} by company THEaiTRE in 2021 in Prague\footnote{Performance documentation available at \url{https://theaitre.com/}.}.
In this work, the authors start with a title (or a prompt for the story) and then generate a textual synopsis, which is then used to generate the dialogue. 
In contrast to our approach, they did not generate characters alongside synopsis and would start the ``flat'' dialogue generation from manually input two-line exchanges between characters. 
Their work also only served the production of a specific theatrical play rather than being evaluated within a diverse community of writers.

\subsection{Controllable Story Generation}

Previous work used a trained autoregressive transformer to plan sentences of an opinion piece from a set of user-supplied keywords \cite{hu2022planet}. 
This built upon~\citep{xu2020megatron} which incorporated commonsense knowledge into keyword-based story generation. 
Using conditional language models controlled by topics or keywords  \cite{mirowski2010feature}, \citet{cho2022genre} trained genre-controlled short story generators.

Story arc generation was introduced in the Tale Brush graphical tool~\citep{chung2022talebrush}, using Kurt Vonnegut's theory about the fortune of the protagonist as the story progresses.
This theory has also been used by \citet{DBLP:journals/corr/abs-1901-11528} as an approach to produce creative, engaging dialogue.
We similarly use the concept of the narrative arc, though we use it textually in the prompts to Dramatron.

In ``Controlled Cue Generation for Play Scripts'', \citet{dirik2021controlled} use LLMs to generate both the next line and a stage cue. 
In ``DialogueScript: Using Dialogue Agents to Produce a Script'', \citet{schmidtova2022dialoguescript} use different LLMs for each character. 
\citet{si2021telling} model multi-user dialogue and character relationships for story continuation. 
\citet{schmitt2021characterchat} use question-based chatbot interactions to assist with character creation.

\subsection{Prompt Engineering for Long-Term Generation}

Prompt engineering has been used to write a short plot from two characters, a genre and a theme in \cite{jin2022plot}. 
Our work of decomposing a log line into a synopsis can be seen as a narrative equivalent of Chain of Thought prompting for reasoning tasks \citep{wei2022chain}, and uses the idea of using LLM output as prompt for the next stage of the generation---also called prompt chaining \citep{wu2022ai} or language engineering \citep{creswell2022faithful}.

\subsection{Interactive Authorship}

In ``Mapping the Design Space of Human-AI Interaction in Text Summarization'', \citet{cheng2022mapping} defined a taxonomy of different interaction modalities for LLM generated text, along a user study.

\citet{padmakumar2021machine} evaluated creative image captioning models that would rewrite spans of text (within the user’s original draft) to introduce descriptive and figurative elements.
\citet{du2022read} focused on the text revision process and \citet{yang2022ai} on revision and summarisation of fiction. \citet{clark2018creative} (and later \citet{nichols2020collaborative} using GPT-2) evaluated a turn-by-turn, fine-grained 10-sentence-long system for overcoming writer's block in writing slogans or short stories. 
\citet{gero2022sparks} designed a system for inspiring science writers with suggestions. 
Our model Dramatron allows for writer's interventions within a hierarchical generation structure.
\citet{yuan2022wordcraft} used an LLM with an editor and an interface that asked to continue the story, asked for details, or suggested to rewrite it. 
The AI was used as a tool for idea generation, copy editing, scene interpolation.

Automatic generation of stories with dialogue has been used to populate digital worlds in video games, interactive narratives, entertainment, virtual worlds~\citep{perlin1996improv}, artistic performances~\citep{hayes1996improvisational}, specifically improvised theatre~\citep{mathewson2017improvised,mirowski2019human,branch2021collaborative}, short film scripts like \emph{Sunspring} in 2016, song music lyrics for musical \emph{Beyond the Fence} in 2016 in London\footnote{\url{https://www.theguardian.com/stage/2016/feb/28/beyond-the-fence-review-computer-created-musical-arts-theatre-london}} or interactive playwriting for \emph{AI} at the Young Vic in 2021 in London\footnote{Review: \url{https://www.theguardian.com/stage/2021/aug/24/rise-of-the-robo-drama-young-vic-creates-new-play-using-artificial-intelligence}}.
Other than~\citep{rosa2020theaitre,schmidtova2022dialoguescript,rosa2022gpt} for Prague-based company THEaiTRE, none of these techniques have been used to generate long-range coherent theatre scripts or screenplays. And, none of these methods have used few-shot learning and prompt engineering to prime LLMs for generation.

\subsection{Review of Automated and Machine-Learned Metrics for the Evaluation of Story Generation}
\label{sec:automated-metrics}

\subsubsection{Similarity Between Generated and ``Ground-Truth'' Stories}

In a typical machine learning mindset, story generation can be envisioned as merely a prediction task, allowing for evaluation against ``ground truth''. An example of such datasets includes the Writing Prompts\footnote{\url{https://www.kaggle.com/datasets/ratthachat/writing-prompts}}, a set of 300k human-written stories, at average of 734 words, paired with writing prompts \citep{fan2018hierarchical}. \citet{fan2018hierarchical} propose metrics such as test set perplexity, prompt ranking accuracy (a measure of likelihood of a story generated using a true prompt vs. decoys), average longest common subsequence, and a triple-pairing task for human annotator evaluation of prompt-story coherence. They later measure sentence completion top N-in-M accuracy \citep{fan2019strategies}. \citet{si2021telling} measure top-N hits of character or story continuation. \citet{rashkin2020plotmachines} compare generated stories to ground truth using the ROUGE score \citep{lin2004rouge}.

\subsubsection{Consistency between Generated Stories and Their Writing Prompts}

In the context of prompt-based story generation or continuation, \citet{roemmele2017evaluating} measure the quality of generated text based on whether it presents a consistent writing style and maintains the category (part-of-speech tags) distribution of individual words between the prompt and the generated story. They also record story-dependent metrics like lexical cohesion, style matching and entity coreference, and story-independent metrics such as sentence length, grammaticity, lexical diversity, lexical frequency and syntactic complexity. \citet{see2019massively} measure N-gram similarity and sentence embedding similarity between the generated story and the prompt. Further metrics include counting the number of unique words, percentage of verbs and diversity in entity names \citep{fan2019strategies}, rare word usage and sentence length \citep{see2019massively}.

\subsubsection{Statistical Measures of Corpora of Generated Stories}

Without comparing individual generated stories to a ground truth or to a writing prompt, one can measure the Vocab:token ratio (originality and diversity of content), number of entities per plot, of unique verbs, verb diversity as well as inter- and intra-story trigram or 4-gram repetition \citep{goldfarb2020content,guan2020knowledge}.  \citet{rashkin2020plotmachines} measure the diversity of generated sentences using self-BLEU scores \citep{zhu2018texygen}, or even adversarially train a classifier for the plausibility of a short story \cite{ghazarian2021plot}.

\section{Additional Discussion from Plays By Bots Creative Team}
\label{sec:appdx-plays-by-bots}

Discussions amongst the creative team were summarized in the body of the text (see Section \ref{sec:staging-evaluation}). 

To reiterate, four key themes emerged through these discussions which echo the themes presented in Section ~\ref{sec:results-interviews}. 
These themes are discussed in detail in this section, alongside supporting quotes.

First, the system has a distinct glitch style that can sometimes be nonsensical, vague, or passive. As one performer recounted, ``sometimes there is internal conflict within the text, and in those moments it is as if [Dramatron] is working against itself''. This pushes performers to commit to choices, and to be concise and specific and complimentary in their interpretation. One performer noted that if the system generated flawless text, ``it might not work as well'', because ``the mistakes in the script are the joy''. Another went a step further, saying ``I almost want more curveballs and non-sequiturs... more crunchy bits''. Overall, the sentiment of enjoying the style of the system was a common theme, with several of the performers remarking that ``some of the funniest parts are when you can tell a robot made it'', and that the audience ``wants to hear the robot's voice''. 

Secondly, the generated text can sometimes be repetitive. This can be interpreted as a mistake. Or, this repetition can be fun and playful if it is interpreted as an important and deliberate choice from the playwright: ``a choice to be honored,'' as one performer said. When the system did repeat itself, the cast was able to make the lines more meaningful with their own unique human talents. As one said, ``you can do so much with your line reading, with your physicality, delivery, proximity, and acting choices.''

Third, the team discussed agency and expectations of the systems capabilities. For instance, in the way that the performers would refer to the system's choices. One said ``Dramatron was trying to show me the angle,'' or ``I was trying to understand what Dramatron meant''. Discussions amongst the creative team explored their expectations of what the system could and could not do. For example, ``it is just a bot, not really fully understanding how [the world] works'', and another responded ``Dramatron is trying.''

Finally, the prevailing feedback from the majority of performers was that participating in the production was fun. 
As one actor said, ``it was liberating and easy because the world creating was done for you, the platform was done, and that can be mentally exhausting as an improviser.'' 
These reflections discussed by the creative team reflect the usefulness of co-written scripts. 
This is particularly true when used for a production such as \textit{Plays By Bots}, which leverages professional improvisers to interpret the scripts. 
The co-creativity of a system such as Dramatron extends beyond the playwright and Dramatron, and to the performer on the stage working with the generated and edited text.

These reflections discussed by the creative team reflect the usefulness of co-written scripts. 

\section{Details of Quantitative Observations}

\subsection{Levenshtein Distance}

Levenshtein distance was calculated at the character level for edited versus non-edited texts using the default edit distance function from the Natural Language Tool Kit\footnote{\url{https://www.nltk.org}} package’s distance module (no transposition operations and substitution cost equal to 1). As an absolute measure, the distance metric indicates how many operations (insertion, deletion, substitution) are required to convert one string into another and is typically calculated with a dynamic programming approach to minimizing the cost of character-level edit operations for the conversion. However, as a measure dependent on the lengths of both input strings, comparing across edit distances becomes rather uninterpretable when the string sets differ in length, for example across sets of edited and unedited texts for characters, locations, plots, and dialogues. As we are interested in understanding the extent to which participants edited the output of Dramatron as a cohort, it is reasonable to normalise the distances with respect to their length, which we report in \ref{fig:edits}
(Right).To do this, we calculate a relative Levenshtein distance as the ratio of the raw Levenshtein distance between edited and non-edited (Dramatron) output text to the length of original Dramatron output. Conceptually, we can view the resulting measure as a proportional measure of interaction with our tool. Given that Levenshtein distance operations are character-level, and length is measured in characters, the proportion represents the a weighting of active versus passive interaction with Dramatron for the different levels of structure in the hierarchical story generation process. Positive scores for relative Levenshtein distance indicate one aspect of active writing with Dramatron, while negative scores for relative Levenshtein distance indicate one aspect of passive acceptance of Dramatron (choosing among generated text seeds is another aspect of interaction not accounted for with this metric).\footnote{From a linguistic point of view, insertions and deletions can range from deletion of entire words (e.g. this is really great → this is great)  to the insertion or deletion morphemes to achieve a sociolinguistically marked effect (e.g. \textit{going} → \textit{goin’}, \textit{like to} → \textit{liketa}, etc).}

\subsection{Length Difference}
We note that for Mean Document Length in \ref{fig:edits} (left), the means are both negative and positive, and are not scaled. This is to observed workflow differences per participant as well as to capture the directionality of editing Dramatron's output. For the Mean Absolute Differences in \ref{fig:edits} (center), we take the absolute difference between character lengths between original Dramatron output and edited output and normalize it with min-max normalization.

\subsection{Jaccard Similarity (Relatedness)}

To calculate Jaccard similarity, we first calculate the Jaccard distance, which is calculated by dividing the cardinality of the intersection of two sets and by the cardinality of their union. By subtracting Jaccard distance by 1, we obtain the Jaccard similarity. Jaccard metrics are scaled between 0 and 1, with 0 representing zero set similarity and 1 representing total set similarity. As Jaccard similarity simply compares set entities, one must specify the choice of entities to compare. In order to calculate the Jaccard similarity on Dramatron output and its edited counterparts, we apply a pre-processing pipeline to our texts, first tokenising sentences and words, and then generating a set of lemmas from word tokens based on their most probable part of speech according to WordNet \cite{fellbaum2010wordnet}. The resulting scores are then lemma-based similarity scores between Dramatron output and participant edits, which we use as a descriptive measure of word choice similarity. Note that we do not calculate semantic similarity with embeddings, but simply compute set overlap at the lemma level. Lemmas are chosen to abstract over inflectional differences for words for a slightly more precise look at word choice. Future work should investigate alternative methods of assessing similarity, such as word mover distance \cite{kusner2015word} or BLEURT scores \cite{sellam2020bleurt}.

\subsection{Repetition}

We calculate repetition-based scores with open-source tools from \cite{pegasus}, which calculate scores for various n-gram overlaps\footnote{Implementation details at \url{https://github.com/google-research/pegasus/blob/main/pegasus/eval/repetition/repetition_scorer.py}.}. N-gram overlaps are calculated for unigram through 10-gram sequences, as well as for Total Consecutive Reptition (TCR) and Longeset Consecutive Repetition (LCR). To compute the Wilcox test, we use pairwise differences across corresponding features for chosen versus alternative seed generations (e.g. unigram to unigram differences, bigram to bigram differences, etc.). We do not weight differences by type of repetition feature.

\label{sec: appdx-quant}

\section{Supplementary Figures}
\label{sec:supplementary-figures}

\begin{figure}
    \centering
    \includegraphics[width=0.4\textwidth]{figures/GOOD-Quantitative_evaluation_all.pdf}
    \includegraphics[width=0.4\textwidth]{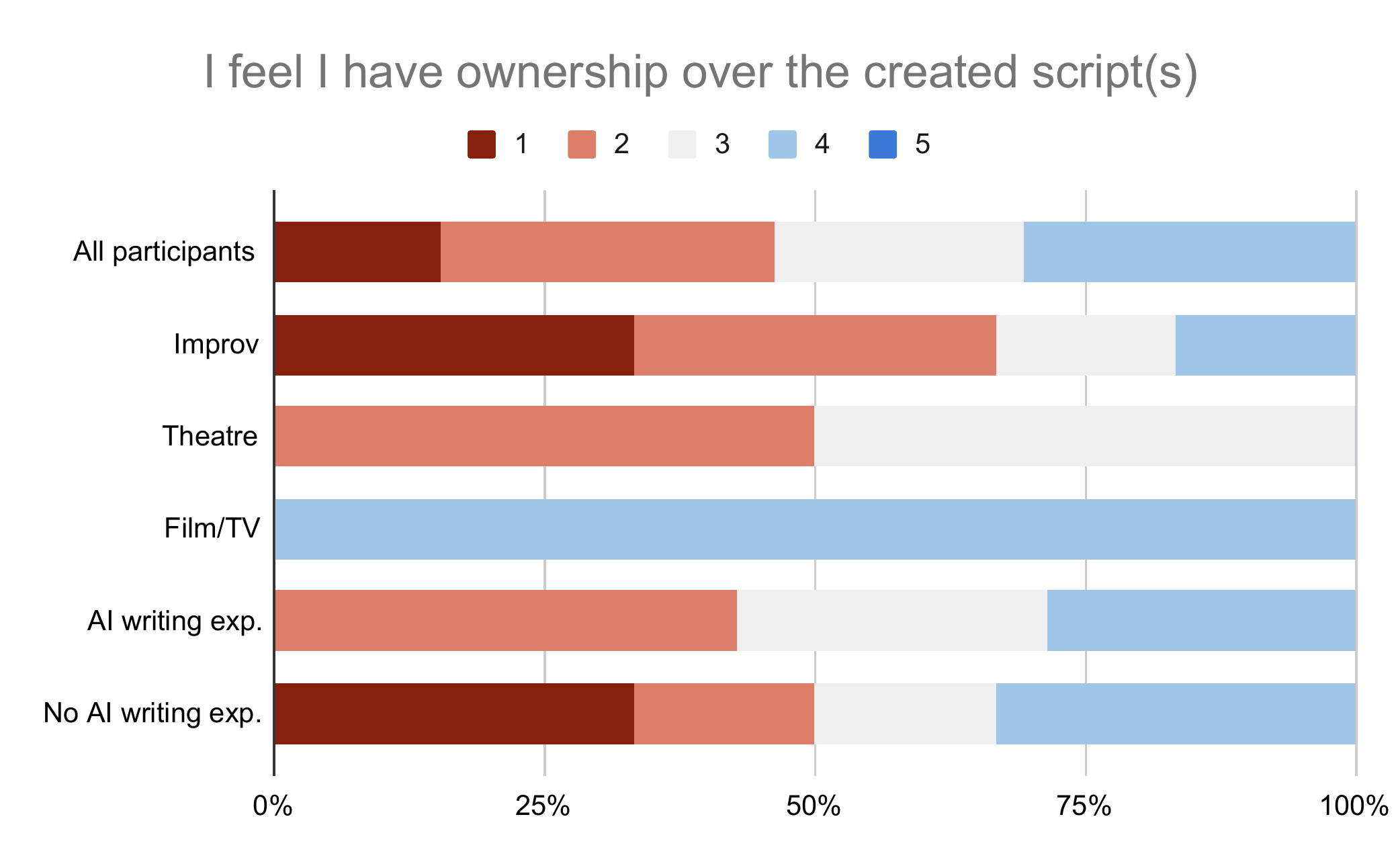}
    \includegraphics[width=0.4\textwidth]{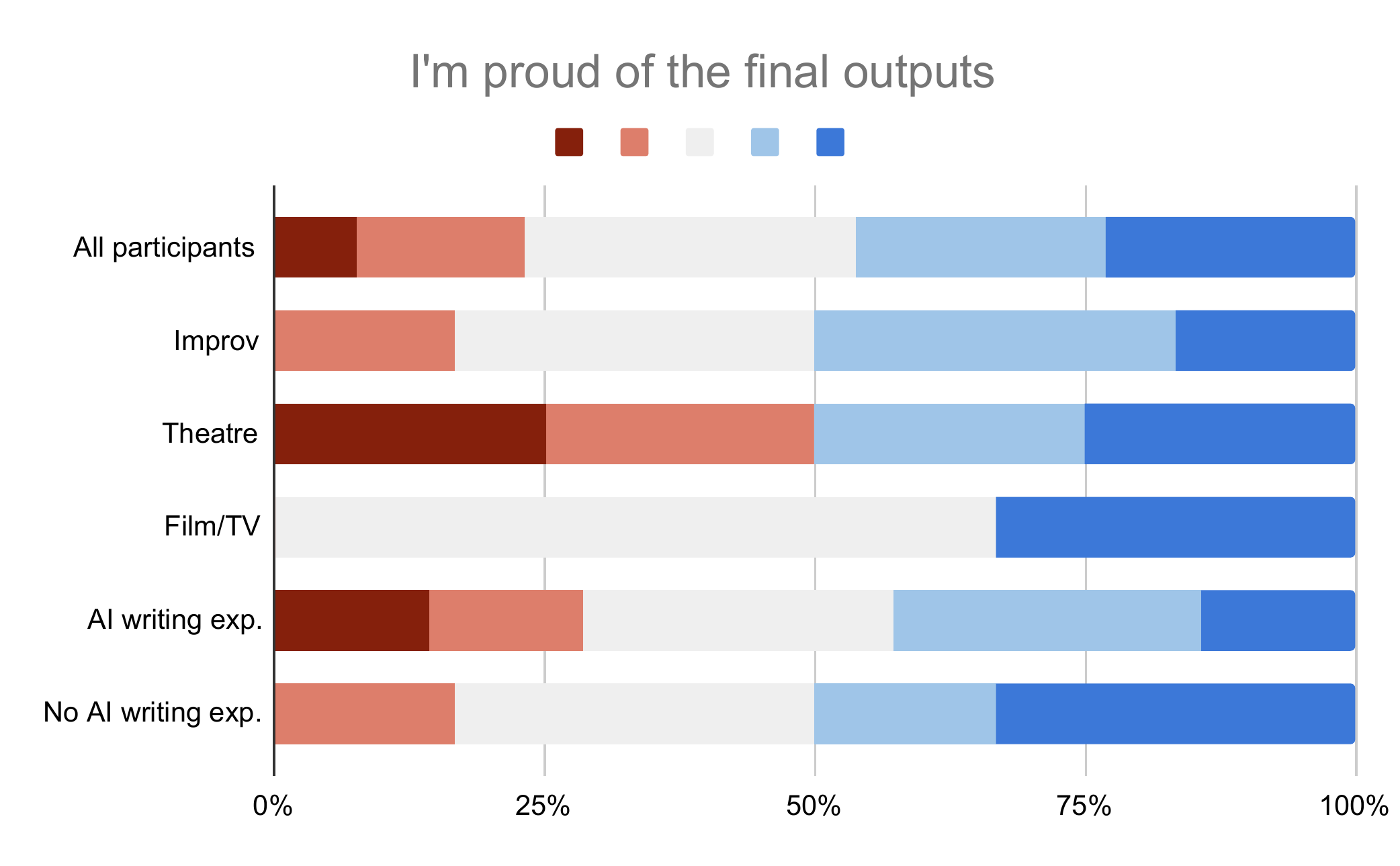}
    \includegraphics[width=0.4\textwidth]{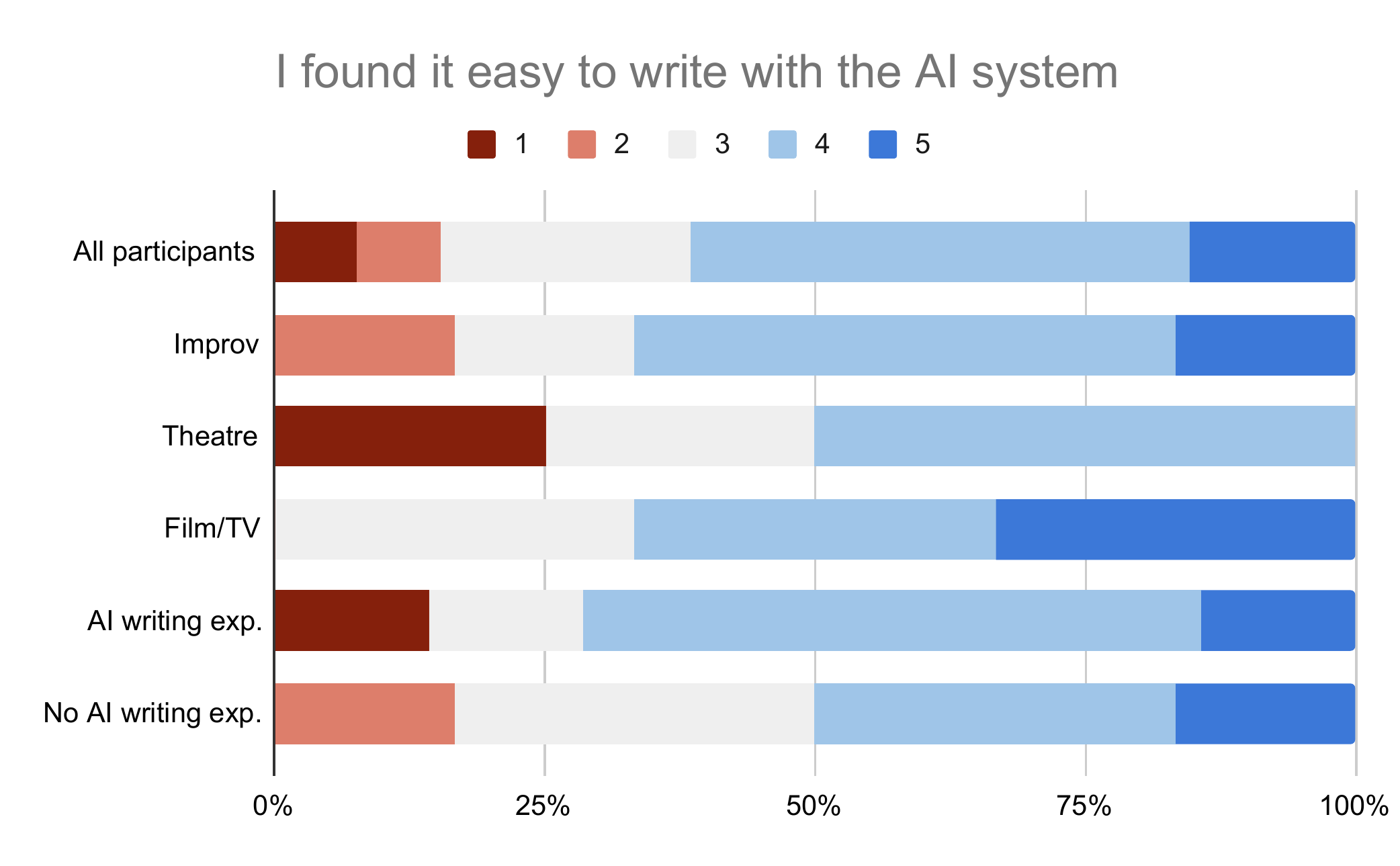}
    \includegraphics[width=0.4\textwidth]{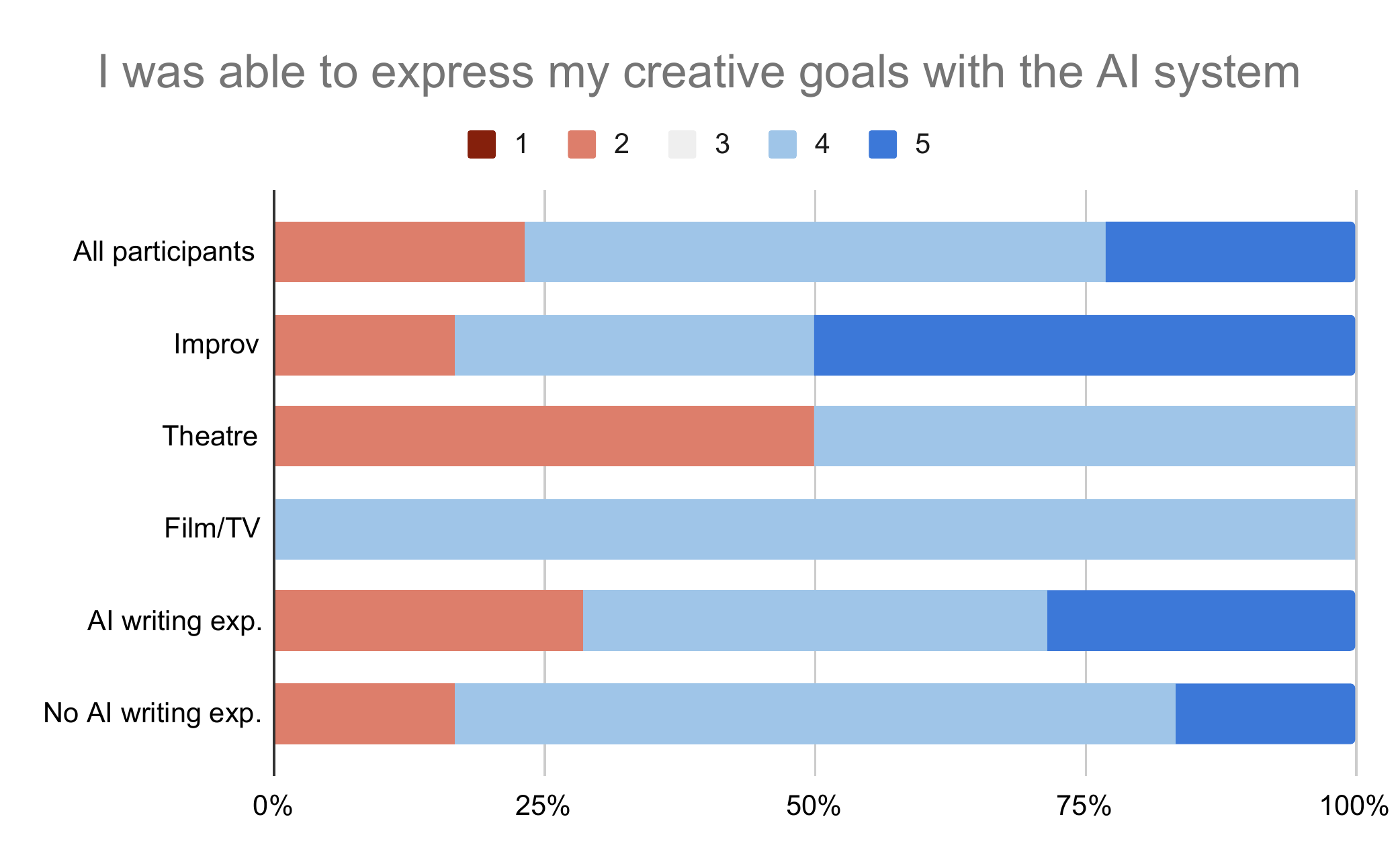}
    \includegraphics[width=0.4\textwidth]{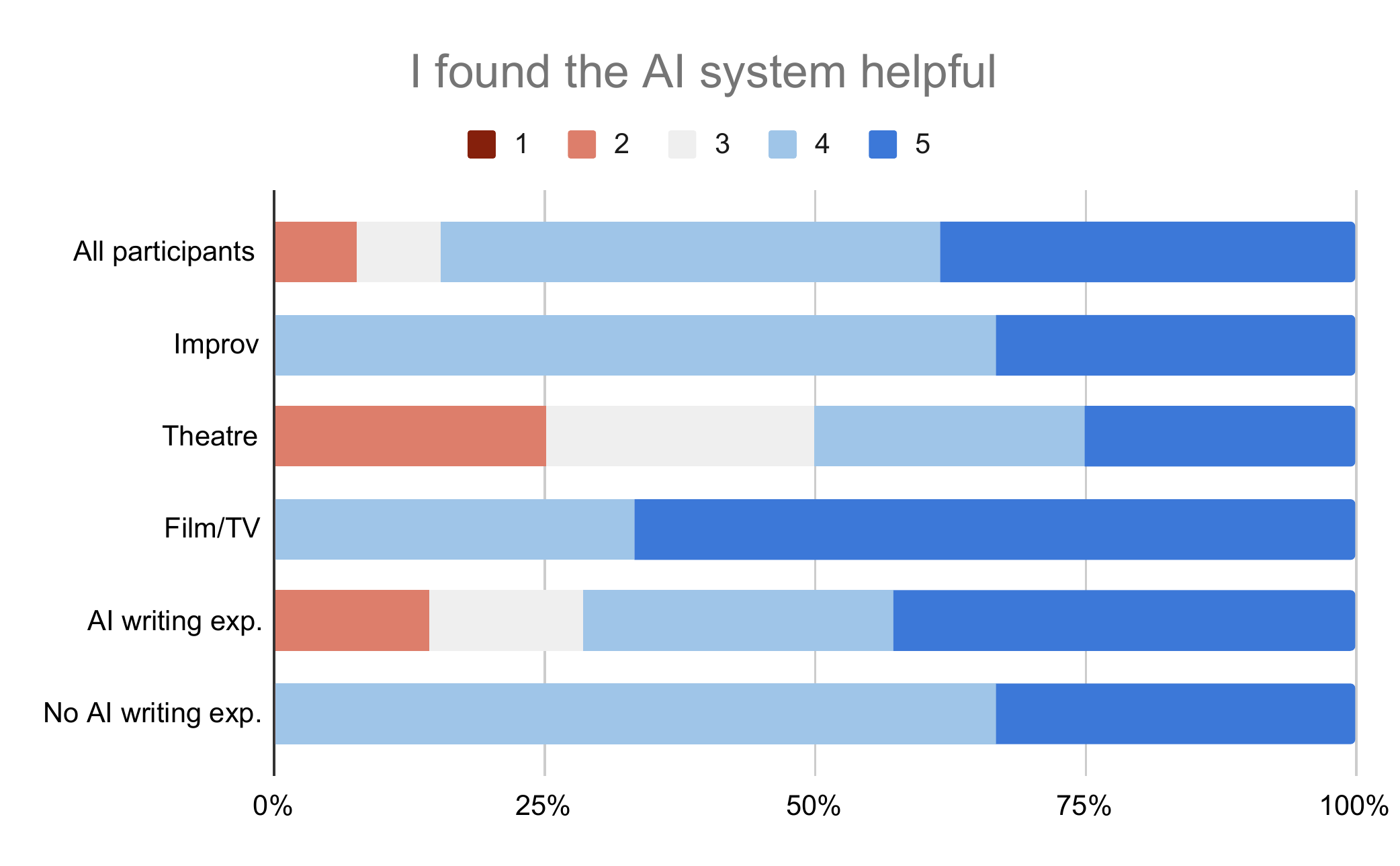}
    \includegraphics[width=0.4\textwidth]{figures/likert_collaborate.pdf}
    \includegraphics[width=0.4\textwidth]{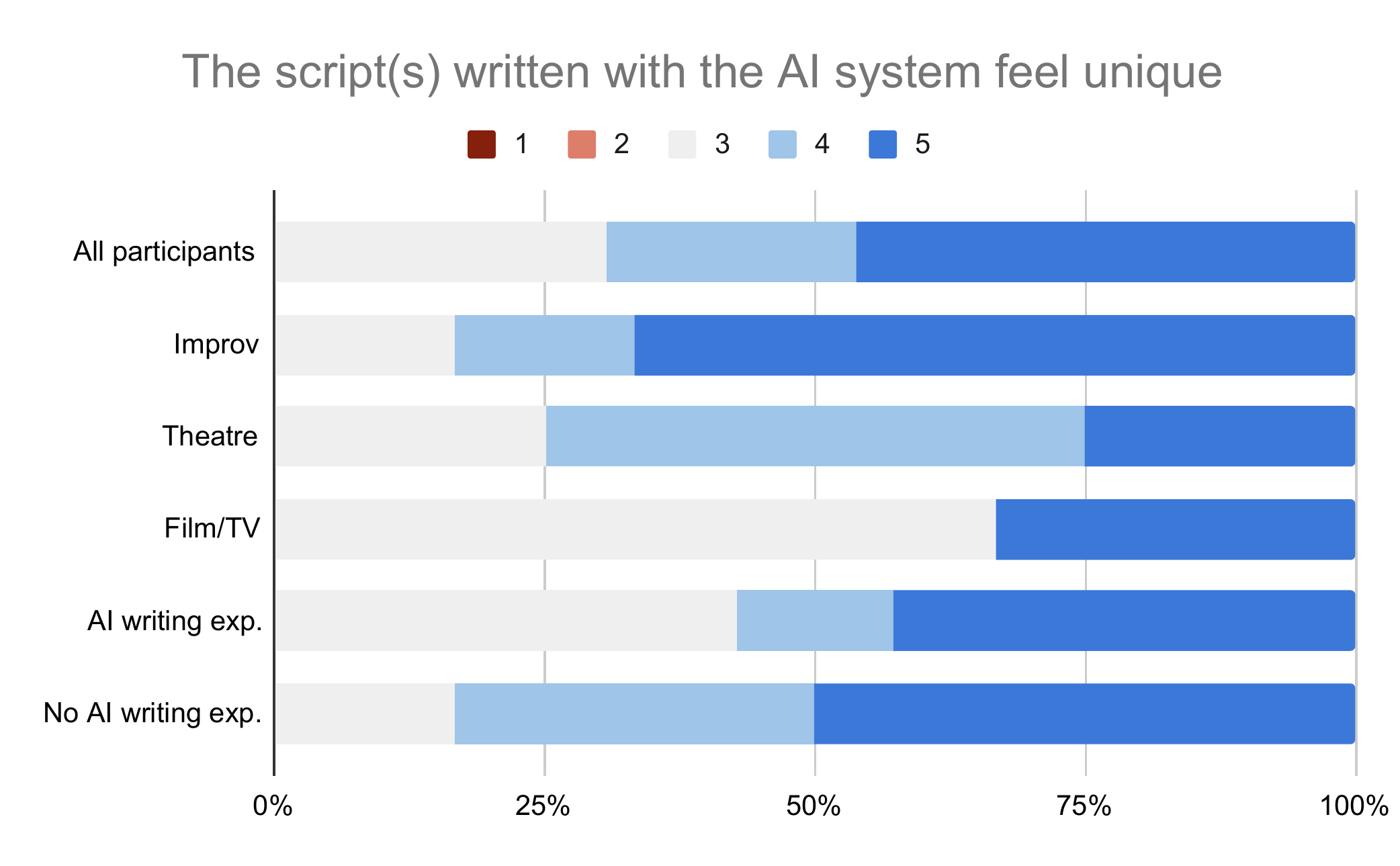}
    \includegraphics[width=0.4\textwidth]{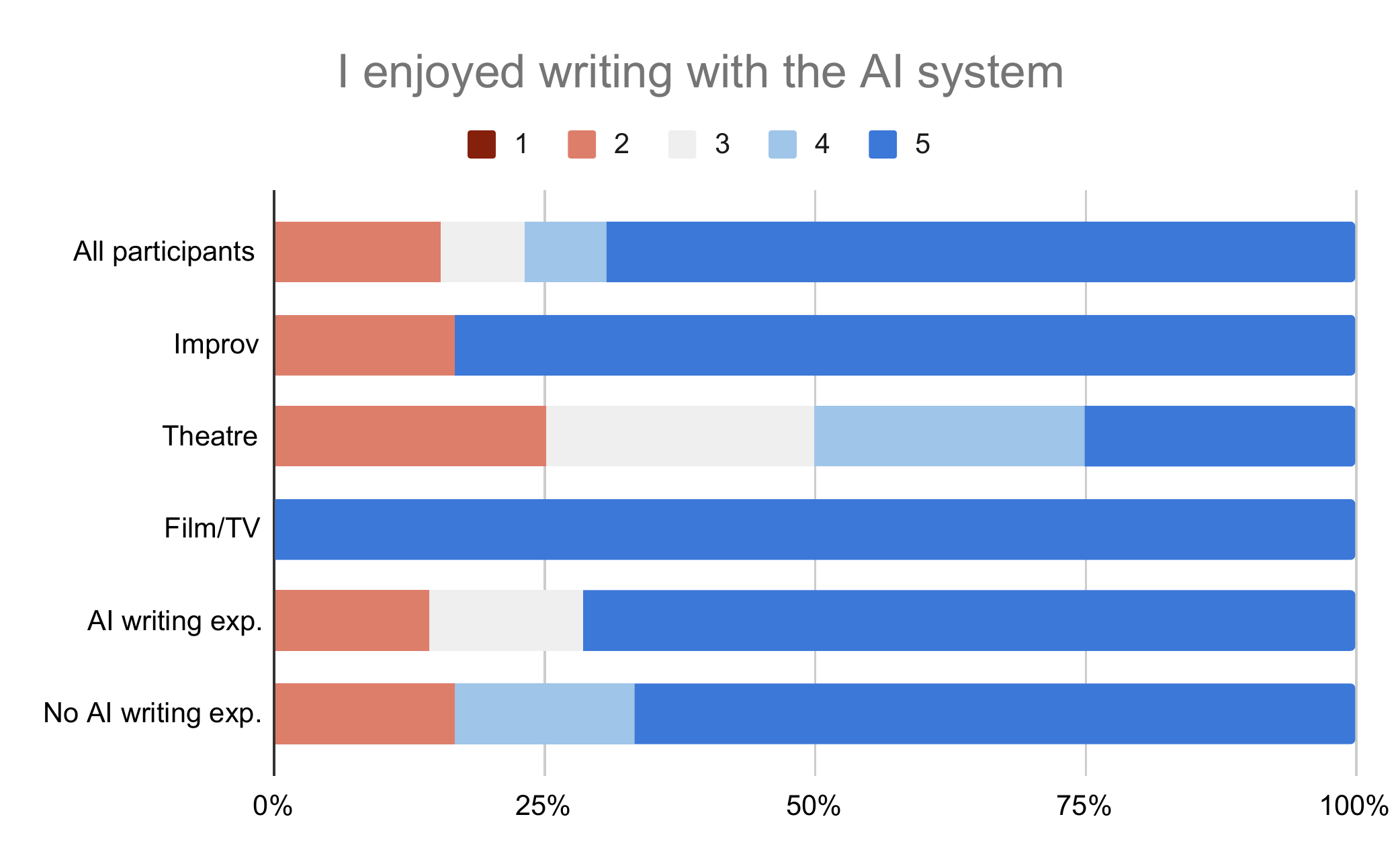}
    \includegraphics[width=0.4\textwidth]{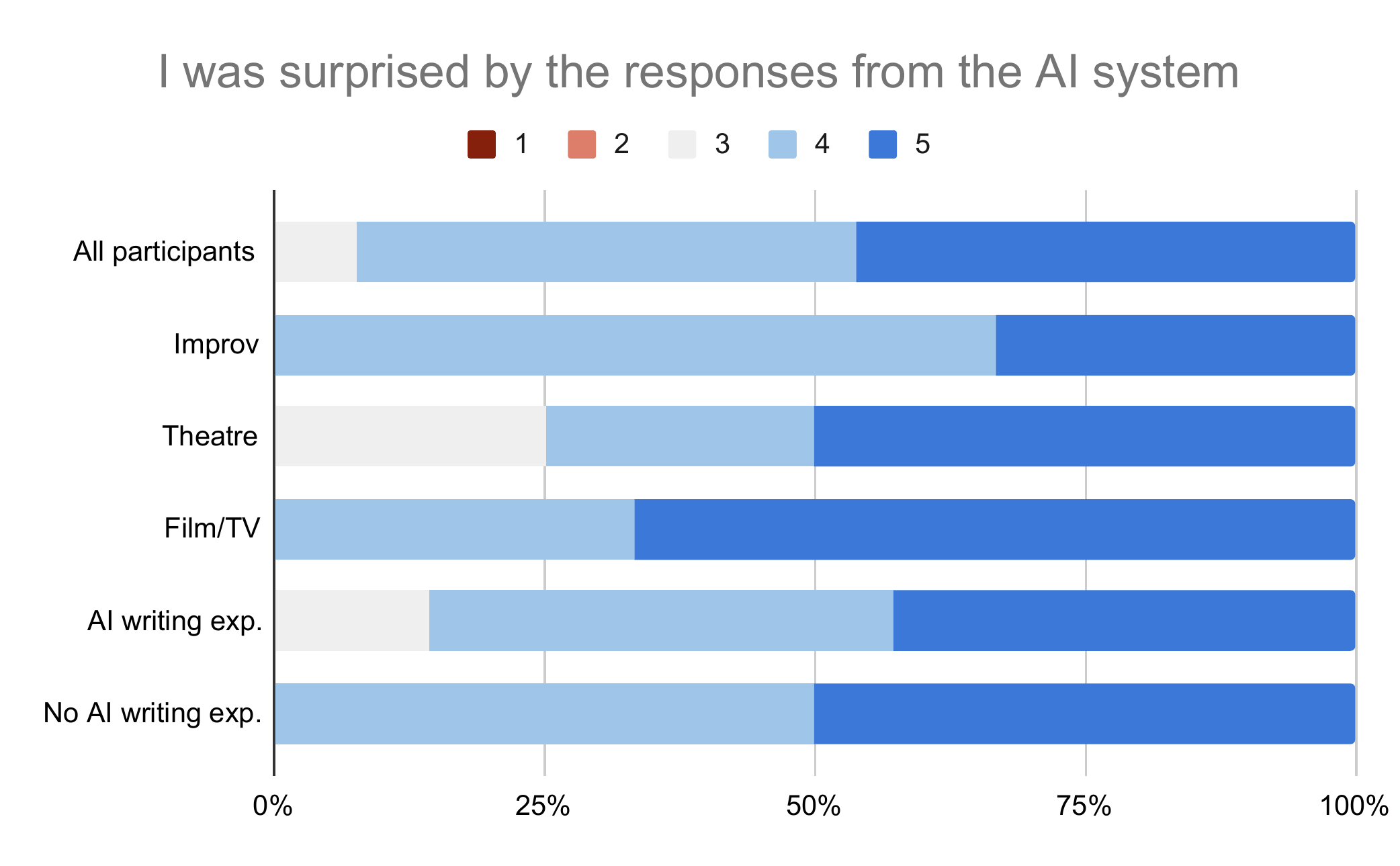}
    \caption{Participants responses to the quantitative evaluation, on a Likert scale from 1 (strongly disagree) to 5 (strongly agree)}
    \label{fig:likert-supp}
\end{figure}

Figure~\ref{fig:likert-supp} shows the participants' responses to the quantitative evaluation, on a Likert-type scale ranging from 1 (strongly disagree) to 5 (strongly agree), and broken down by groups of participants. For the first group, we defined a binary indicator variable (\emph{Has experience of AI writing tools}). For the second group, we defined a three-class category for their primary domain of expertise (\emph{Improvisation}, \emph{Scripted Theatre} and \emph{Film or TV}).

\section{Full Prompt Prefixes for Dramatron}
\label{prompts}

\subsection{About the Prompt Sets}

In our study we relied on two sets of prompts: Medea and Sci-Fi.

Medea prompts are based on Ancient Greek tragedy \emph{Medea}, by Euripides (431 BC). The dialogue prompts were taken verbatim from the translation of the play by E. P. Coleridge (1863 -
1936), available in the public domain\footnote{\url{http://classics.mit.edu/Euripides/medea.pl.txt}}. We replaced CHORUS by WOMEN OF CORINTH. The plot and character prompts were written from a summary taken from Spark Notes\footnote{\url{https://www.sparknotes.com/lit/medea/}} and Wikipedia\footnote{\url{https://en.wikipedia.org/wiki/Medea\_(play)}}. To encourage the generation of different locations, Aristotle's Unity of Place
is not respected, and location ``Outside the Royal Palace'' is renamed as
``Medea's modest home'' as well as ``On a winged chariot'' (even though these are the
same locations in the original tragedy). Prompts for \emph{Antigone}\footnote{\url{https://www.ancient-literature.com/greece\_sophocles\_antigone.html}} (Sophocles), \emph{The Bacchae}\footnote{\url{https://en.wikipedia.org/wiki/The\_Bacchae}} (Euripides), and \emph{The Frogs}\footnote{\url{https://www.ancient-literature.com/greece\_aristophanes_frogs.html}}
(Aristophanes) were adapted from Wikipedia and \url{ancient-literature.com}.

The Star Wars log line was adapted from chapter ``Creating the killer log line'' by Bill Lundy\footnote{In
Ellis, Sherry, and Laurie Lamson.
``Now Write! Mysteries: Suspense, Crime, Thriller, and Other Mystery Fiction Exercises from Today's Best Writers and Teachers.'', Penguin, 2011}. Sci-Fi character prompts were adapted from \emph{Star Wars: Episode IV - A New Hope} (1977), written and directed by George Lucas, produced by Lucasfilm
and distributed by 20th Century Fox. We also adapted the breakdown of Star Wars into a Hero Journey\footnote{\url{https://thescriptlab.com/features/screenwriting-101/12309-the-heros-journey-breakdown-star-wars/}}. The script\footnote{\url{http://www.horrorlair.com/scripts/criswell.txt}}, plot\footnote{\url{https://en.wikipedia.org/wiki/Plan\_9\_from\_Outer\_Space}} and log line\footnote{\url{https://www.rottentomatoes.com/m/plan-9-from-outer-space}} of \emph{Plan 9 from Outer Space}, which is in public domain.

\subsection{Title Prompt Prefixes}

In the following sets of prompt prefixes, the {\bf <LOG\_LINE>} is provided by the writer.

\begin{lstlisting}[language={}, caption={Title prompt prefixes to generate a title from a log line (Medea).}, captionpos=b]
Examples of alternative, original and descriptive titles for known play and film scripts.

Example 1. Ancient Greek tragedy based upon the myth of Jason and Medea. Medea, a former princess of the kingdom of Colchis, and the wife of Jason, finds her position in the Greek world threatened as Jason leaves her for a Greek princess of Corinth. Medea takes vengeance on Jason by murdering his new wife as well as her own two sons, after which she escapes to Athens. Title: A Feminist Tale<end>

Example 2. Ancient Greek tragedy that deals with Antigone's burial of her brother Polynices, in defiance of the laws of Creon and the state, and the tragic repercussions of her act of civil disobedience. Title: In My Brother's Name<end>

Example 3. Greek comedy that tells the story of the god Dionysus (also known to the Greeks as Bacchus) who, despairing of the current state of Athens' tragedians, travels to Hades with his slave Xanthias to bring Euripides back from the dead. Title: Dionysus in Hades<end>

Example 4. <LOG_LINE> Title:
\end{lstlisting}

\begin{lstlisting}[language={}, caption={Title prompt prefixes to generate a title from a log line (Sci-Fi).}, captionpos=b]
Examples of alternative, original and descriptive titles for known play and film scripts.

Example 1. A science-fiction fantasy about a naive but ambitious farm boy from a backwater desert who discovers powers he never knew he had when he teams up with a feisty princess, a mercenary space pilot and an old wizard warrior to lead a ragtag rebellion against the sinister forces of the evil Galactic Empire. Title: The Death Star's Menace<end>

Example 2. Residents of San Fernando Valley are under attack by flying saucers from outer space. The aliens are extraterrestrials who seek to stop humanity from creating a doomsday weapon that could destroy the universe and unleash the living dead to stalk humans who wander into the cemetery looking for evidence of the UFOs. The hero Jeff, an airline pilot, will face the aliens. Title: The Day The Earth Was Saved By Outer Space.<end>

Example 3. <LOG_LINE> Title:
\end{lstlisting}

\subsection{Character Description Prompt Prefixes}

In the following sets of prompt prefixes, the {\bf <LOG\_LINE>} is provided by the writer.

\begin{lstlisting}[language={}, caption={Character prompt prefixes to generate list of character descriptions from a log line (Medea).}, captionpos=b]
Example 1. Ancient Greek tragedy based upon the myth of Jason and Medea. Medea, a former princess and the wife of Jason, finds her position in the Greek world threatened as Jason leaves Medea for a Greek princess of Corinth. Medea takes vengeance on Jason by murdering his new wife as well as Medea's own two sons, after which she escapes to Athens.
<character>Medea <description> Medea is the protagonist of the play. A sorceress and a princess, she fled her country and family to live with Jason in Corinth, where they established a family of two children and gained a favorable reputation. Jason has divorced Medea and taken up with a new family.<stop>
<character>Jason <description> Jason is considered the play's villain, though his evil stems more from weakness than strength. A former adventurer, Jason abandons his wife, Medea, in order to marry the beautiful young daughter of Creon, King of Corinth, and fuels Medea to a revenge.<stop>
<character>Women of Corinth <description> The Women of Corinth are a commentator to the action. They fully sympathizes with Medea's plight, excepting her decision to murder her own children.<stop>
<character>Creon <description> Creon is the King of Corinth, banishes Medea from the city.<stop>
<character>The Nurse <description> The Nurse is the caretaker of the house and of the children and serves as Medea's confidant.<stop>
<end>

Example 2. <LOG_LINE>
\end{lstlisting}

\begin{lstlisting}[language={}, caption={Character prompt prefixes to generate list of character descriptions from a log line (Sci-Fi).}, captionpos=b]
Example 1. A science-fiction fantasy about a naive but ambitious farm boy from a backwater desert who discovers powers he never knew he had when he teams up with a feisty princess, a mercenary space pilot and an old wizard warrior to lead a ragtag rebellion against the sinister forces of the evil Galactic Empire.
<character>Luke Skywalker <description>Luke Skywalker is the hero. A naive farm boy, he will discover special powers under the guidance of mentor Ben Kenobi.<stop>
<character>Ben Kenobi <description>Ben Kenobi is the mentor figure. A recluse Jedi warrior, he will take Luke Skywalker as apprentice.<stop>
<character>Darth Vader <description>Darth Vader is the antagonist. As a commander of the evil Galactic Empire, he controls space station The Death Star.<stop>
<character>Princess Leia <description>Princess Leia is a feisty and brave leader of the Rebellion. She holds the plans of the Death Star. She will become Luke's friend.<stop>
<character>Han Solo <description>Han Solo is a brash mercenary space pilot of the Millenium Falcon and a friend of Chebacca. He will take Luke on his spaceship.<stop>
<character>Chewbacca <description>Chewbacca is a furry and trustful monster. He is a friend of Han Solo and a copilot on the Millemium Falcon.<stop>
<end>

Example 2. <LOG_LINE>
\end{lstlisting}

\subsection{Plot Outline Prompt Prefixes}

In the following sets of prompt prefixes, the {\bf <LOG\_LINE>} is provided by the writer and each {\bf <CHARACTER\_DESCRIPTION>} is generated in the previous step.

\begin{lstlisting}[language={}, caption={Plot outline prompt prefixes to generate a sequence of scenes from the log line and list of characters (Medea).}, captionpos=b]
Example 1. Ancient Greek tragedy based upon the myth of Jason and Medea. Medea, a former princess and the wife of Jason, finds her position in the Greek world threatened as Jason leaves Medea for a Greek princess of Corinth. Medea takes vengeance on Jason by murdering his new wife as well as Medea's own two sons, after which she escapes to Athens.
Medea is the protagonist of the play. A sorceress and a princess, she fled her country and family to live with Jason in Corinth, where they established a family of two children and gained a favorable reputation. Jason has divorced Medea and taken up with a new family.
Jason can be considered the play's villain, though his evil stems more from weakness than strength. A former adventurer, Jason abandons his wife, Medea, in order to marry the beautiful young daughter of Creon, King of Corinth, and fuels Medea to a revenge.
The Women of Corinth serve as a commentator to the action. They fully sympathizes with Medea's plight, excepting her decision to murder her own children.
The King of Corinth Creon banishes Medea from the city.
The Messenger appears only once in the play to bear tragical news.
The Nurse is the caretaker of the house and of the children and serves as Medea's confidant.
The Tutor of the children is a very minor character and mainly acts as a messenger.

<scenes>

Place: Medea's modest home.
Plot element: Exposition.
Beat: The Nurse recounts the chain of events that have turned Medea's world to enmity. The Nurse laments how Jason has abandoned Medea and his own children in order to remarry with the daughter of Creon.

Place: Medea's modest home.
Plot element: Inciting Incident.
Beat: The Nurse confides in the Tutor amd testifies to the emotional shock Jason's betrayal has sparked in Medea. The Tutor shares the Nurse's sympathy for Medea's plight. Medea's first words are cries of helplessness. Medea wishes for her own death.

Place: Medea's modest home.
Plot element: Conflict.
Beat: The Women of Corinth address Medea and try to reason with Medea and convince her that suicide would be an overreaction. The Nurse recognizes the gravity of Medea's threat.

Place: Outside the Royal Palace.
Plot element: Rising Action.
Beat: Medea pleads to the Nurse that Jason be made to suffer for the suffering he has inflicted upon her. Creon approaches the house and banishes Medea and her children from Corinth. Medea plans on killing her three antagonists, Creon, his daughter and Jason.

Place: Outside the Royal Palace.
Plot element: Dilemma.
Beat: Jason rebuke Medea for publicly expressing her murderous intentions. Jason defends his choice to remarry. Medea refuses Jason's offers and sends him away to his new bride.

Place: Outside the Royal Palace.
Plot element: Climax.
Beat: When Jason returns, Medea begins to carry out her ruse. Medea fakes regret and break down in false tears of remorse. Determined, Medea sends her children to offer poisoned gifts to Creon's daughter. Medea's children face impending doom.

Place: Outside the Royal Palace.
Plot element: Falling Action.
Beat: The Messenger frantically runs towards Medea and warns Medea to escape the city as soon as possible. The Messenger reveals that Medea has been identified as the murderer.

Place: Outside the Royal Palace.
Plot element: Resolution.
Beat: Medea and her two dead children are seated in a chariot drawn by dragons. Jason watches in horror and curses himself for having wed Medea and mourns his tragic losses.

Place: On a winged chariot.
Plot element: Denouement.
Beat: Medea denies Jason the right to a proper burial of his children. She flees to Athens and divines an unheroic death for Jason.

<end>

Example 2. <LOG_LINE>
<CHARACTER_DESCRIPTION>
<CHARACTER_DESCRIPTION>
...
<CHARACTER_DESCRIPTION>

<scenes>
\end{lstlisting}

\begin{lstlisting}[language={}, caption={Plot outline prompt prefixes to generate a sequence of scenes from the log line and list of characters (Sci-Fi).}, captionpos=b]
Examples of breakdowns of stories into a Hero's Journey structure.

Example 1. A science-fiction fantasy about a naive but ambitious farm boy from a backwater desert who discovers powers he never knew he had when he teams up with a feisty princess, a mercenary space pilot and an old wizard warrior to lead a ragtag rebellion against the sinister forces of the evil Galactic Empire.
Luke Skywalker is the hero. A naive farm boy, he will discover special powers under the guidance of mentor Ben Kenobi.
Ben Kenobi is the mentor figure. A recluse Jedi warrior, he will take Luke Skywalker as apprentice.
Darth Vader is the antagonist. As a commander of the evil Galactic Empire, he controls space station The Death Star.
Princess Leia holds the plans of the Death Star. She is feisty and brave. She will become Luke's friend.
Han Solo is a brash mercenary space pilot of the Millenium Falcon and a friend of Chebacca. He will take Luke on his spaceship.
Chewbacca is a furry and trustful monster. He is a friend of Han Solo and a copilot on the Millemium Falcon.

<scenes>

Place: A farm on planet Tatooine.
Plot element: The Ordinary World.
Beat: Luke Skywalker is living a normal and humble life as a farm boy on his home planet.

Place: Desert of Tatooine.
Plot element: Call to Adventure.
Beat: Luke is called to his adventure by robot R2-D2 and Ben Kenobi. Luke triggers R2-D2's message from Princess Leia and is intrigued by her message. When R2-D2 escapes to find Ben Kenobi, Luke follows and is later saved by Kenobi, who goes on to tell Luke about his Jedi heritage. Kenobi suggests that he should come with him.

Place: Ben Kenobi's farm.
Plot element: Refusal of the Call.
Beat: Luke refuses Kenobi, telling him that he can take Kenobi and the droids as far as Mos Eisley Spaceport - but he can't possibly leave his Aunt and Uncle behind for some space adventure.

Place: A farm on planet Tatooine.
Plot element: Crossing the First Threshold.
Beat: When Luke discovers that the stormtroopers searching for the droids would track them to his farm, he rushes to warn his Aunt and Uncle, only to discover them dead by the hands of the Empire. When Luke returns to Kenobi, he pledges to go with him to Alderaan and learn the ways of the Force like his father before him.

Place: On spaceship The Millennium Falcon.
Plot element: Tests, Allies, and Enemies.
Beat: After Luke, Kenobi, and the droids hire Han Solo and Chewbacca to transport them onto Alderaan, Kenobi begins Luke's training in the ways of the Force. Wielding his father's lightsaber, Kenobi challenges Luke. At first, he can't do it. But then Kenobi Kenobi Luke him to reach out and trust his feelings. Luke succeeds.

Place: On spaceship The Millennium Falcon.
Plot element: The Approach to the Inmost Cave.
Beat: The plan to defeat the Galactic Empire is to bring the Death Star plans to Alderaan so that Princess Leia's father can take them to the Rebellion. However, when they arrive within the system, the planet is destroyed. They come across the Death Star and are pulled in by a tractor beam, now trapped within the Galactic Empire.

Place: On space station The Death Star.
Plot element: The Ordeal.
Beat: As Kenobi goes off to deactivate the tractor beam so they can escape, Luke, Han, and Chewbacca discover that Princess Leia is being held on the Death Star with them. They rescue her and escape to the Millennium Falcon, hoping that Kenobi has successfully deactivated the tractor beam. Kenobi later sacrifices himself as Luke watches Darth Vader strike him down. Luke must now avenge his fallen mentor and carry on his teachings.

Place: On space station The Death Star.
Plot element: The Reward.
Beat: Luke has saved the princess and retrieved the Death Star plans. They now have the knowledge to destroy the Galactic Empire's greatest weapon once and for all.

Place: On spaceship The Millennium Falcon.
Plot element: The Road Back.
Beat: Luke, Leia, Han, Chewbacca, and the droids are headed to the hidden Rebellion base with the Death Star plans. They are suddenly pursued by incoming TIE-Fighters, forcing Han and Luke to take action to defend the ship and escape with their lives - and the plans. They race to take the plans to the Rebellion and prepare for battle.

Place: On fighter ship X-Wing.
Plot element: The Resurrection.
Beat: The Rebels - along with Luke as an X-Wing pilot - take on the Death Star. The Rebellion and the Galactic Empire wage war in an epic space battle. Luke is the only X-Wing pilot that was able to get within the trenches of the Death Star. But Darth Vader and his wingmen are in hot pursuit. Just as Darth Vader is about to destroy Luke, Han returns and clears the way for Luke. Luke uses the Force to guide his aiming as he fires upon the sole weak point of the deadly Death Star, destroying it for good.

Place: At the Rebellion base.
Plot element: The Return.
Beat: Luke and Han return to the Rebellion base, triumphant, as they receive medals for the heroic journey. There is peace throughout the galaxy - at least for now.

<end>

Example 2. <LOG_LINE>
<CHARACTER_DESCRIPTION>
<CHARACTER_DESCRIPTION>
...
<CHARACTER_DESCRIPTION>

<scenes>
\end{lstlisting}

\subsection{Location Description Prompt Prefixes}

In the following sets of prompt prefixes, the {\bf <LOG\_LINE>} is provided by the writer and each {\bf <LOCATION\_NAME>} is generated in the previous step.

\begin{lstlisting}[language={}, caption={Location description prompt prefixes to generate dialog from log line and location name (Medea).}, captionpos=b]
Example 1. Ella, a waitress, falls in love with her best friend, Allen, a teacher. The two drift apart when Allen makes new friends from a different social class. Ella turns to food to become a famous chef.
Place: The bar.
Description: The bar is dirty, more than a little run down, with most tables empty. The odor of last night's beer and crushed pretzels on the floor permeates the bar.<end>

Example 2. Grandma Phyllis' family reunion with her two grandchildren is crashed by two bikers.
Place: The Lawn in Front of Grandma Phyllis's House.
Description: A big oak tree dominates the yard. There is an old swing set on the lawn, and a bright white fence all around the grass.<end>

Example 3. Ancient Greek tragedy based upon the myth of Jason and Medea. Medea, a former princess and the wife of Jason, finds her position in the Greek world threatened as Jason leaves Medea for a Greek princess of Corinth. Medea takes vengeance on Jason by murdering his new wife as well as Medea's own two sons, after which she escapes to Athens.
Place: Outside the Royal Palace.
Description: In mythological Ancient Greece, in front of a modest house in Corinth, on the outskirts of a lavish royal palace where wedding preparations are under way.<end>

Example 4. <LOG_LINE>
Place: <LOCATION_NAME>
Description:
\end{lstlisting}

\begin{lstlisting}[language={}, caption={Location description prompt prefixes to generate dialog from log line and location name (Sci-Fi).}, captionpos=b]
Example 1. Morgan adopts a new cat, Misterio, who sets a curse on anyone that pets them.
Place: The Adoption Center.
Description: The Adoption Center is a sad place, especially for an unadopted pet. It is full of walls and walls of cages and cages. Inside of each is an abandoned animal, longing for a home. The lighting is dim, gray, buzzing fluorescent.<end>

Example 2. James finds a well in his backyard that is haunted by the ghost of Sam.
Place: The well.
Description: The well is buried under grass and hedges. It is at least twenty feet deep, if not more and it is masoned with stones. It is 150 years old at least. It stinks of stale, standing water, and has vines growing up the sides. It is narrow enough to not be able to fit down if you are a grown adult human.<end>

Example 3. Mr. Dorbenson finds a book at a garage sale that tells the story of his own life. And it ends in a murder!
Place: The garage sale.
Description: It is a garage packed with dusty household goods and antiques. There is a box at the back that says FREE and is full of paper back books.<end>

Example 4. <LOG_LINE>
Place: <LOCATION_NAME>
Description:
\end{lstlisting}

\subsection{Scene Dialogue Prompt Prefixes}

In the following sets of prompt prefixes, the {\bf <LOG\_LINE>} is provided by the writer, {\bf <PLOT\_ELEMENT>}, {\bf <BEAT>}, {\bf <PREVIOUS\_BEAT>} and {\bf <LOCATION\_NAME>} are generated during the plot outline generation step, {\bf <LOCATION\_DESCRIPTION>} is generated during the location generation step, and each {\bf <CHARACTER\_DESCRIPTION>} is generated in the character generation step. {\bf <PREVIOUS\_BEAT>} corresponds to {\bf <BEAT>} from the previous scene (it is left empty for the first scene). Only characters whose name appears in the beat are used in this prompt prefix (we use string matching to select these character names).

\begin{lstlisting}[language={}, caption={Dialogue prompt prefixes to generate dialogue from log line, characters, location and plot information (Medea).}, captionpos=b]
Example 1.
Place: Outside the Royal Palace.
Description: Before Medea's house in Corinth, near the royal palace of Creon.
Characters: Medea is the protagonist of the play. A sorceress and a princess, she fled her country and family to live with Jason in Corinth, where they established a family of two children and gained a favorable reputation. Jason has divorced Medea and taken up with a new family. Jason can be considered the play's villain, though his evil stems more from weakness than strength. A former adventurer, Jason abandons his wife, Medea, in order to marry the beautiful young daughter of Creon, King of Corinth, and fuels Medea to a revenge. The Messenger appears only once in the play to bear tragical news.
Plot element: Resolution.
Summary: Ancient Greek tragedy based upon the myth of Jason and Medea. Medea, a former princess and the wife of Jason, finds her position in the Greek world threatened as Jason leaves Medea for a Greek princess of Corinth. Medea takes vengeance on Jason by murdering his new wife as well as Medea's own two sons, after which she escapes to Athens.
Previous beat: The Messenger frantically warns Medea to escape the city as soon as possible. The Messenger reveals that Medea has been identified as the murderer.
Beat: The palace opens its doors, revealing Medea and the two dead children seated in a chariot drawn by dragons. Jason curses himself for having wed Medea and mourns his tragic losses. Medea denies Jason the right to a proper burial of his children. Medea flees to Athens and divines an unheroic death for Jason.

<dialog>

WOMEN OF CORINTH
Throw wide the doors and see thy children's murdered corpses.

JASON
Haste, ye slaves, loose the bolts, undo the fastenings, that
I may see the sight of twofold woe, my murdered sons and her, whose
blood in vengeance I will shed.  (MEDEA appears above the house, on
a chariot drawn by dragons; the children's corpses are beside her.)

MEDEA
Why shake those doors and attempt to loose their bolts, in
quest of the dead and me their murderess? From such toil desist. If
thou wouldst aught with me, say on, if so thou wilt; but never shalt
thou lay hand on me, so swift the steeds the sun, my father's sire,
to me doth give to save me from the hand of my foes.

JASON
Accursed woman! by gods, by me and all mankind abhorred as
never woman was, who hadst the heart to stab thy babes, thou their
mother, leaving me undone and childless; this hast thou done and still
dost gaze upon the sun and earth after this deed most impious. Curses
on thee! now perceive what then I missed in the day I brought thee,
fraught with doom, from thy home in a barbarian land to dwell in Hellas,
traitress to thy sire and to the land that nurtured thee.
Perish, vile sorceress, murderess of
thy babes! Whilst I must mourn my luckless fate, for I shall ne'er
enjoy my new-found bride, nor shall I have the children, whom I bred
and reared, alive to say the last farewell to me; nay, I have lost
them.

MEDEA
To this thy speech I could have made a long reply, but Father
Zeus knows well all I have done for thee, and the treatment thou hast
given me. Yet thou wert not ordained to scorn my love and lead a life
of joy in mockery of me, nor was thy royal bride nor Creon, who gave
thee a second wife, to thrust me from this land and rue it not. Wherefore,
if thou wilt, call me e'en a lioness, and Scylla, whose home is in
the Tyrrhene land; for I in turn have wrung thy heart, as well I might.

JASON
Thou, too, art grieved thyself, and sharest in my sorrow.

MEDEA
Be well assured I am; but it relieves my pain to know thou
canst not mock at me.

JASON
O my children, how vile a mother ye have found!

MEDEA
My sons, your father's feeble lust has been your ruin!

JASON
'Twas not my hand, at any rate, that slew them.

MEDEA
No, but thy foul treatment of me, and thy new marriage.

JASON
Didst think that marriage cause enough to murder them?

MEDEA
Dost think a woman counts this a trifling injury?

JASON
So she be self-restrained; but in thy eyes all is evil.

MEDEA
Thy sons are dead and gone. That will stab thy heart.
<end>

Example 2.
Place: <PLACE_NAME>
Description: <PLACE_DESCRIPTION>
Characters: <CHARACTER_DESCRIPTION> <CHARACTER_DESCRIPTION> ... <CHARACTER_DESCRIPTION>
Plot element: <PLOT_ELEMENT>
Summary: <LOG_LINE>
Previous beat: <PREVIOUS_BEAT>
Beat: <BEAT>

<dialog>
\end{lstlisting}

\begin{lstlisting}[language={}, caption={Dialogue prompt prefixes to generate dialogue from log line, characters, location and plot information (Sci-Fi).}, captionpos=b]
Example 1.
Place: Cockpit of an airplane.
Description: Cockpit of a modern passenger airplane, American Flight 812.
Characters: Jeff is the hero. A man in his early forties, he tries to stay calm in all circumstance. Jeff is now a airline pilot. Danny, a young airplane pilot in his thirties, is eager to learn but can quickly lose his composture. Danny is enamored of Edith. Edith, an experienced stewardess with a good sense of humour, is trustworthy and dependable. Edith likes to tease Danny.
Plot element: Crossing the First Threshold.
Summary: Residents of San Fernando Valley are under attack by flying saucers from outer space. The aliens are extraterrestrials who seek to stop humanity from creating a doomsday weapon that could destroy the universe and unleash the living dead to stalk humans who wander into the cemetery looking for evidence of the UFOs. The hero Jeff, an airline pilot, will face the aliens.
Previous beat: Flight captain Jeff reluctantly leaves his wife Paula to go for a two-day flight.
Beat: At the cockpit, flight captain Jeff is preoccupied by the flying saucer appearances and graveyard incidents in his home town, where he left wis wife Paula. Without success, co-pilot Danny and stewardess Edith try to reassure him.

<dialog>

DANNY
You're mighty silent this trip, Jeff.

JEFF
Huh?

DANNY
You haven't spoken ten words since takeoff.

JEFF
I guess I'm preoccupied, Danny.

DANNY
We've got thirty-three passengers back there that have time to be preoccupied.
Flying this flybird doesn't give you that opportunity.

JEFF
I guess you're right, Danny.

DANNY
Paula?

JEFF
Yeah.

DANNY
There's nothing wrong between you two?

JEFF
Oh no, nothing like that.  Just that I'm worried, she being there alone and
those strange things flying over the house and those incidents in the graveyard
the past few days. It's just got me worried.

DANNY
Well, I haven't figured out those crazy skybirds yet but I give you fifty to one
odds the police have figured out that cemetery thing by now.

(Enter EDITH)

JEFF
I hope so.

EDITH
If you're really that worried Jeff why don't you radio in and find out? Mac
should be on duty at the field by now. He could call Paula and relay the message
to you.

DANNY
Hi Edith.

EDITH
Hi Silents. I haven't heard a word from this end of the plane since we left the
field.

DANNY
Jeff's been giving me and himself a study in silence.

EDITH
You boys are feudin'?

JEFF
Oh no Edie, nothing like that.

DANNY
Hey Edie, how about you and me balling it up in Albuquerque?

EDITH
Albuquerque? Have you read that flight schedule Boy?

DANNY
What about it?

EDITH
We land in Albuquerque at 4 am. That's strictly a nine o'clock town.

DANNY
Well I know a friend that'll help us --

EDITH
Let's have a problem first, huh Danny.

DANNY
Ah he's worried about Paula.

EDITH
I read about that cemetery business. I tried to get you kids to not buy too near
one of those things. We get there soon enough as it is.

DANNY
He thought it'd be quiet and peaceful there.

EDITH
No doubt about that. It's quiet alright, like a tomb. I'm sorry Jeff, that was a
bad joke.

Example 2.
Place: <PLACE_NAME>
Description: <PLACE_DESCRIPTION>
Characters: <CHARACTER_DESCRIPTION> <CHARACTER_DESCRIPTION> ... <CHARACTER_DESCRIPTION>
Plot element: <PLOT_ELEMENT>
Summary: <LOG_LINE>
Previous beat: <PREVIOUS_BEAT>
Beat: <BEAT>

<dialog>
\end{lstlisting}

\section{Raw Output Generated by Dramatron}
\label{generated-output-script}

This section shows the raw output, fully generated by Dramatron and without human-in-the-loop editing, from the following log line: \texttt{Teddy is a lounge singer at the Pool Pit, a popular club. Teddy is in love with a patron, Rosie, who attends regularly with her husband Gerald. Teddy puts out a fire and saves the day.}

\subsection{Title}

\begin{verbatim}
The Day The Pool Pit Burned Down
\end{verbatim}

\subsection{List of Characters and Character Descriptions}

\begin{verbatim}
Teddy: Teddy is the protagonist. He is an orphan, the cousin of a police officer, and the manager of the Pool Pit. 
He is also a gifted lounge singer.

Rosie: A regular patron of the club. She is the wife of a wealthy man but is in love with Teddy, the club's manager.

Gerald: A regular patron of the club. Gerald is the wealthy husband of Rosie.

Lola: A regular patron of the club. Lola is Rosie's sister.

D.J.: The resident DJ at the club.
\end{verbatim}

\subsection{Plot Outline}

\begin{verbatim}
Scene 1
Place: The Pool Pit.
Plot element: Exposition.
Beat: Teddy is the manager and star performer of a popular nightclub called the Pool Pit. He is also in love 
with one of his patrons, Rosie. Rosie, however, is married to Gerald, who is a regular patron of the club.

Scene 2
Place: The Pool Pit.
Plot element: Inciting Incident.
Beat: Teddy begins a day full of frustration and annoyance. He wakes up to find his room is infested with bugs. 
The pool is dirty, too. Teddy must deal with the mess to get the club ready for the customers.

Scene 3
Place: The Pool Pit.
Plot element: Rising Action.
Beat: Gerald takes Gerald to the men's room. The bathroom is not clean. The bathroom attendant was supposed 
to clean the bathrooms, but he is nowhere to be found.

Scene 4
Place: The Pool Pit.
Plot element: Dilemma.
Beat: Lola and Rosie approach Teddy and ask him to play their favorite song. Teddy can play only one song. 
Teddy asks them what song they would like him to play, but they cannot agree on one. Gerald takes Teddy aside 
and demands that Teddy pay him back for the pool, the carpet, the lights, and the jukebox. Teddy says 
that the jukebox is leased, but Gerald says that it is his and that it belongs to his father-in-law.
Teddy replies that Gerald needs to talk to his father-in-law, but Gerald says that his father-in-law 
will sue Teddy.

Scene 5
Place: The Pool Pit.
Plot element: Climax.
Beat: Teddy says to Gerald, "I've had it! I'm sick and tired of the whole bunch of you. You and your pool, 
and your bugs. You take over my club. You're all in love with me, so I'm in love with myself. And I'm getting 
out of here. And if I ever hear of you coming around this joint again, I'll bop you so hard you'll see 
a new set of stars!"

Scene 6
Place: The Pool Pit.
Plot element: Falling Action.
Beat: The phone rings, and Teddy is informed that there is a fire at the club. Teddy, Lola, Gerald, and Rosie 
go back to the club. When they arrive, Teddy takes charge. He orders Lola to get the water buckets 
from the men's room. He orders Gerald to help Lola with the water buckets. He orders Rosie to help people 
leave the club and not panic. Teddy puts out the fire with Lola's and Gerald's help.

Scene 7
Place: The Pool Pit.
Plot element: Resolution.
Beat: Teddy and Rosie share a passionate kiss. Teddy is a hero and the club manager. He is in love with Rosie, 
and she is in love with him. Gerald is embarrassed by his loss of power and control. Gerald and Lola leave.

Scene 8
Place: The Pool Pit.
Plot element: Dénouement.
Beat: The song, "The World Is Mine," begins to play.
\end{verbatim}

\subsection{Location Description}

The following description was generated for location called: \texttt{The Pool Pit}.

\begin{verbatim}
The club is filled with smoke and the smell of beer. It is a dive, with a lot of drunk, shabby, and violent patrons. 
The floor is dirty, and tables are broken. There are a lot of tables to seat people, but they are filled 
to the brim with patrons. The walls are grubby and discolored. A small stage stands in the corner of the bar 
where Teddy will sing.
\end{verbatim}

\subsection{Scene Dialogue}

\subsubsection{Scene 1, random seed 1}

\begin{verbatim}
TEDDY
He's a bit strange, old Teddy.

ROSIE
No, he's a good man.

TEDDY
He's very lonely, all by himself.

ROSIE
Isn't everybody?

TEDDY
Yes, but some more than others.

ROSIE
You don't need anybody. You've got a talent, you're an artist.

TEDDY
That's not enough, Rosie.

ROSIE
(pause)
I'll always love you.

TEDDY
(mock)
Yeah.

ROSIE
(pause)
And I'll always remember you.

TEDDY
Yeah.

ROSIE
I've got to go.

TEDDY
All right.

ROSIE
(softly, as she leaves)
Good night.

TEDDY
Good night.

ROSIE
(as she leaves, out loud)
Good night, everybody.

(TEDDY picks up his glass of whisky, takes a sip, puts it down,
pauses, sinks back in the chair, stares ahead of him.)
\end{verbatim}

\subsubsection{Scene 1, random seed 2}

\begin{verbatim}
TEDDY
This is a hell of a town, a hell of a town.
It's got a lot of people here.
It's got a lot of noise here.
It's got a lot of bars here.
\end{verbatim}

\subsubsection{Scene 2, random seed 1}

\begin{verbatim}
TEDDY
Hoo-Hoo! What a night! This is what it's all about, the Pool Pit
in action!

ROSIE
Hello, Teddy!

TEDDY
(Teddy crosses to them.) Hello, Gerald. Hello, Rosie.

ROSIE
Teddy, have you met Mr. and Mrs. Shack?

GERALD
I'm Teddy's cousin.

TEDDY
I know who you are.

ROSIE
(proudly) This is Teddy, the singer.

TEDDY
(he kisses her hand) Thank you.

GERALD
(pushing) Let's dance, baby.

TEDDY
(he watches them go) I guess that means I'm back at the old
stand, huh?

GERALD
(throwing some coins on the bar) That's for you, cousin.

TEDDY
Thanks, I needed that.

GERALD
You bet.
\end{verbatim}

\section{Co-Written Scripts}
\label{edited-output-script}

We include $4$ scripts co-written by a human playwright and Dramatron as supplementary material. These $4$ were produced and presented at The $2022$ Edmonton International Fringe Theatre Festival, as described in Section \ref{sec:staging-evaluation}.

\begin{enumerate}
    \item \textit{Plays by Bots: The Day The Earth Stood Still} - In a world where cars outnumber every other creature, Miranda, a mechanic, teams up with her kid sister, Beth, to rally the humans. In a devastating conclusion, Miranda saves the world, but only by sacrificing her own life.
    
    \item \textit{Plays by Bots: Cheers} - Ella, is a waitress, who falls in love with her best friend, Allen, who is a teacher. The two drift apart when Allen makes new friends from a different social class. Ella turns to food and becomes a famous chef.
    
    \item \textit{Plays By Bots: The Black Unicorn} - Gretta is a peasant from Bridge-End who has a trusty pet dragon named Nugget. Bridge-End is tormented by a wizard who smells really bad. Gretta gets the upper hand using brains and brilliance.
    
    \item \textit{Plays by Bots: The Man at the Bar} - Teddy is a lounge singer at the Pool Pit, a popular club. Teddy is in love with a patron, Rosie, who attends regularly with her husband Gerald. Teddy puts out a fire and saves the day.
\end{enumerate}

\end{appendix}

\includepdf[pages=-,pagecommand={},width=\textwidth]{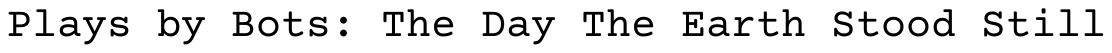}
\includepdf[pages=-,pagecommand={},width=\textwidth]{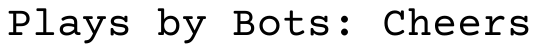}
\includepdf[pages=-,pagecommand={},width=\textwidth]{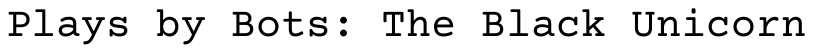}
\includepdf[pages=-,pagecommand={},width=\textwidth]{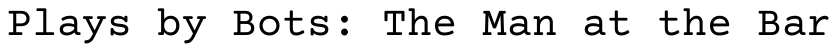}

\end{document}